\documentclass[twocolumn]{aastex61}

\usepackage{graphicx}	
\usepackage[fleqn]{amsmath}	
\usepackage{amssymb}	
\usepackage{exscale}  
\usepackage{relsize}
\usepackage{xcolor}
\usepackage{longtable}
\PassOptionsToPackage{hyphens}{url}\usepackage{hyperref}  
\hypersetup{
	colorlinks   = true, 
	urlcolor     = blue, 
	linkcolor    = blue, 
	citecolor   = blue 
}

\usepackage{float}
\usepackage{footmisc} 
\usepackage{verbatim}

\newcommand{\mch}{M$_{\rm Ch}$}

\newcommand{\crcode}{\texttt{ChN}}

\received{\today}
\revised{}
\accepted{}
\submitjournal{\apj}

\shorttitle{Bulk properties of Type Ia SNRs}
\shortauthors{H. Mart\'{i}nez-Rodr\'{i}guez et al.}

\begin{document}

\title{Chandrasekhar and sub-Chandrasekhar models for the X-ray emission of Type Ia supernova remnants (I): Bulk properties}

\author[0000-0002-1919-228X]{H\'{e}ctor Mart\'{i}nez-Rodr\'{i}guez}
\affiliation{Department of Physics and Astronomy and Pittsburgh Particle Physics, Astrophysics and Cosmology Center (PITT PACC), University of Pittsburgh, 3941 O'Hara Street, Pittsburgh, PA 15260, USA}
\correspondingauthor{H\'{e}ctor Mart\'{i}nez-Rodr\'{i}guez}
\email{hector.mr@pitt.edu}

\author[0000-0003-3494-343X]{Carles Badenes}
\affiliation{Department of Physics and Astronomy and Pittsburgh Particle Physics, Astrophysics and Cosmology Center (PITT PACC), University of Pittsburgh, 3941 O'Hara Street, Pittsburgh, PA 15260, USA}
\affiliation{Institut de Ci\`encies del Cosmos (ICCUB), Universitat de Barcelona (IEEC-UB), Mart\'i Franqu\'es 1, E08028 Barcelona, Spain}

\author[0000-0002-2899-4241]{Shiu-Hang Lee}
\affiliation{Department of Astronomy, Kyoto University, Kyoto 606-8502, Japan}
\affiliation{RIKEN, Astrophysical Big Bang Laboratory, 2-1 Hirosawa, Wako, Saitama 351-0198, Japan}

\author[0000-0002-7507-8115]{Daniel J. Patnaude}
\affiliation{Smithsonian Astrophysical Observatory, 60 Garden St, Cambridge, MA 02138}

\author[0000-0003-3462-8886]{Adam R. Foster}
\affiliation{Smithsonian Astrophysical Observatory, 60 Garden St, Cambridge, MA 02138}

\author[0000-0002-5092-6085]{Hiroya Yamaguchi}
\affiliation{NASA Goddard Space Flight Center, Code 662, Greenbelt, MD 20771, USA}

\author[0000-0002-4449-9152]{Katie Auchettl}
\affiliation{Center for Cosmology and Astro-Particle Physics, The Ohio State University, 191 West Woodruff Avenue, Columbus, OH 43210, USA}
\affiliation{Department of Physics, The Ohio State University, 191 West Woodruff Avenue, Columbus, OH 43210, USA}

\author[0000-0003-0894-6450]{Eduardo Bravo}
\affiliation{E.T.S. Arquitectura del Vall\`{e}s, Universitat Polit\`{e}cnica de Catalunya, Carrer Pere Serra 1-15, E-08173 Sant Cugat del Vall\`{e}s, Spain}

\author[0000-0002-6986-6756]{Patrick O. Slane}
\affiliation{Smithsonian Astrophysical Observatory, 60 Garden St, Cambridge, MA 02138}

\author[0000-0001-6806-0673]{Anthony L. Piro}
\affiliation{Carnegie Observatories, 813 Santa Barbara Street, Pasadena, CA 91101, USA}

\author[ 0000-0003-3900-7739]{Sangwook Park}
\affiliation{Department of Physics, University of Texas at Arlington, Box 19059, Arlington, TX 76019, USA}

\author[0000-0002-7025-284X]{Shigehiro Nagataki}
\affiliation{RIKEN, Astrophysical Big Bang Laboratory, 2-1 Hirosawa, Wako, Saitama 351-0198, Japan}

\begin{abstract}
	
Type Ia supernovae originate from the explosion of carbon-oxygen white dwarfs in binary systems, 
but the exact nature of their progenitors remains elusive. The bulk properties of Type Ia supernova 
remnants, such as the radius and the centroid energy of the Fe K$\alpha$ blend in the X-ray spectrum, 
are determined by the properties of the supernova ejecta and the ambient medium. 
We model the interaction between Chandrasekhar and sub-Chandrasekhar models for Type Ia supernova 
ejecta and a range of uniform ambient medium densities in one dimension up to an age of 5000 years.
We generate synthetic X-ray spectra from these supernova remnant models and compare their bulk 
properties at different expansion ages with X-ray observations from \textit{Chandra} and \textit{Suzaku}. 
We find that our models can successfully reproduce the bulk properties of most observed remnants, 
suggesting that Type Ia SN progenitors do not modify their surroundings significantly on scales of 
a few pc, although more detailed models are required to establish quantitative limits on the 
density of any such surrounding circumstellar material. 
Ambient medium density and expansion age are the main contributors to the diversity of 
the bulk properties in our models. Chandrasekhar and sub-Chandrasekhar progenitors make similar 
predictions for the bulk remnant properties, but detailed fits to X-ray spectra have the power to 
discriminate explosion energetics and progenitor scenarios.

\end{abstract}


\section{Introduction}\label{sec:intro}

Type Ia supernovae (SNe Ia) are the thermonuclear explosions of 
white dwarf (WD) stars that are destabilized by mass accretion from a close 
binary companion. They are important for a wide range of topics in astrophysics,
e.g. galactic chemical evolution \citep{Ko06, An16,Pr18}, studies of dark energy 
\citep{Ri98,Pe99} and constraints on $\Lambda$CDM parameters \citep{Be14,Res14}. 
Yet, basic aspects of SN Ia physics, such as 
the nature of their stellar progenitors and the 
triggering mechanism for the thermonuclear runaway, still remain obscure. 
Most proposed scenarios for the progenitor systems of SNe Ia fall into two broad categories: 
the single degenerate (SD), where the WD companion
is a nondegenerate star, and the double degenerate (DD), where the WD companion
is another WD \citep[see] [for recent reviews]{Wa12,Ma14,Liv18,Sok18,Wa18}.

In the SD scenario, the WD accretes material from its companion over a relatively 
long timescale (t$ \, {\sim} \, 10^{6}$ years) and explodes 
when its mass approaches the Chandrasekhar limit 
M$_{\rm{Ch}} \, {\simeq} \, 1.4 \, M_{\odot}$ \citep{No84,Th86,Ha96,Ha04}. 
Conversely, in most DD scenarios, the WD becomes unstable after a merger or a collision
 on a dynamical timescale \citep{Ib84} and explodes with a mass
 that is not necessarily close to \mch $\,$ 
\citep[e.g.,][]{Ras09,vaK10,Ku13}. In theory, distinguishing between SD 
and DD systems should be feasible, given that some observational probes 
are sensitive to the duration of the accretion process or to the total mass prior to the explosion
 \citep[e.g.,][]{Ba07,Ba08a,Sei13a,Mrg14,Sca14,Ya15,Chom16,MR16}.

Sub-Chandrasekhar models \citep[e.g.,][]{Wo94,Si10,Wo11} are a 
particular subset of both SD and DD SN Ia progenitors. To first order, 
the mass of $\rm{^{56}Ni}$ synthesized, and therefore the brightness of the supernova, 
is determined by the mass of the exploding WD.
A sub-\mch\ WD cannot detonate spontaneously without some kind of external compression --  
double-detonations are frequently invoked \citep[e.g.,][]{Sh13,Sh14a,Sh14b,Sh18}. 
Here, a carbon-oxygen (C/O) WD accretes material from a companion 
and develops a helium-rich layer that eventually becomes unstable,
ignites, and sends a shock wave into the core. This blast wave converges and
creates another shock that triggers a carbon denotation, which 
explodes the WD. Violent mergers \citep[e.g.,][]{Pak12,Pak13} are an alternative
scenario where, right before the secondary WD is disrupted, carbon burning
starts on the surface of the primary WD and a detonation propagates through the
whole merger, triggering a thermonuclear runaway. 
Other studies present pure detonations of sub-\mch\ C/O 
WDs with different masses without addressing the question of how they were 
initiated. However, these studies are still able to reproduce many 
observables such as light curves, nickel ejecta masses, and isotopic mass 
ratios \citep{Si10,Pi14,Ya15,Blo17,MR17,Gol18,Sh18}.

After the light from the supernova (SN) fades away, the
ejecta expand and cool down until their density becomes comparable to that of 
the ambient medium, either the interstellar medium (ISM) 
or a more or less extended circumstellar medium (CSM) modified by the 
SN progenitor. At this point, the supernova remnant (SNR) phase begins. 
The ejecta drive a blast wave into the ambient medium 
(``forward shock'', FS), and the pressure gradient 
creates another wave back into the ejecta 
\citep[``reverse shock'', RS;][]{McK95,Tru99}.

\begin{table*}
\footnotesize
\begin{center}
\caption{ Total yields for the sub-\mch\ and \mch\ progenitor models. See \citet{Br18} for details and  extended yields. \label{table:progenitor_models}}
\hskip-3cm
\makebox[1 \textwidth][c]{
\begin{tabular}{ccccccccccccc}
\tableline
\noalign{\smallskip}
\tableline
\noalign{\smallskip}
Progenitor & $M_{\rm{C}}$  & $M_{\rm{O}}$ & $M_{\rm{Ne}}$ & $M_{\rm{Mg}}$ & $M_{\rm{Si}}$ & $M_{\rm{S}}$ & $M_{\rm{Ar}}$ & $M_{\rm{Ca}}$ & $M_{\rm{Cr}}$ & $M_{\rm{Mn}}$ & $M_{\rm{Fe}}$ & $M_{\rm{Ni}}$  \\
\noalign{\smallskip}
$\mathrm{}$ & $(M_{\odot})$ & $(M_{\odot})$ & $(M_{\odot})$ & $(M_{\odot})$ & $(M_{\odot})$ & $(M_{\odot})$ & $(M_{\odot})$ & $(M_{\odot})$ & $(M_{\odot})$ & $(M_{\odot})$ & $(M_{\odot})$ & $(M_{\odot})$ \\
\noalign{\smallskip}
\tableline
\noalign{\smallskip}
SCH088 & 3.95E-03 & 1.40E-01 & 2.54E-03 & 1.99E-02 & 2.79E-01 & 1.66E-01 & 3.70E-02 & 3.72E-02 & 6.90E-03 & 2.68E-03 & 1.82E-01 & 1.19E-03 \\
SCH097 & 1.62E-03 & 7.66E-02 & 8.24E-04 & 7.80E-03 & 2.09E-01 & 1.36E-01 & 3.26E-02 & 3.52E-02 & 1.12E-02 & 4.24E-03 & 4.50E-01 & 3.18E-03 \\
SCH106 & 6.91E-04 & 3.74E-02 & 2.80E-04 & 2.61E-03 & 1.38E-01 & 9.62E-02 & 2.39E-02 & 2.63E-02 & 9.11E-03 & 3.46E-03  & 7.01E-01 & 1.54E-02 \\
SCH115 & 2.75E-04 & 1.47E-02 & 8.99E-05 & 6.34E-04 & 7.66E-02 & 5.66E-02 & 1.47E-02 & 1.66E-02 & 6.31E-03 & 2.40E-03 & 9.25E-01 & 2.71E-02 \\
\noalign{\smallskip}
\tableline
\noalign{\smallskip}
DDT12 & 4.88E-03 & 1.75E-01 & 3.88E-03 & 2.64E-02 & 3.84E-01 & 2.34E-01 & 5.29E-02 & 5.32E-02 & 1.50E-02 & 7.12E-03 & 3.84E-01 & 3.15E-02 \\
DDT16 & 2.52E-03 & 1.19E-01 & 1.83E-03 & 1.55E-02 & 3.05E-01 & 1.98E-01 & 4.79E-02 & 5.20E-02 & 2.02E-02 & 8.76E-03 & 5.70E-01 & 3.16E-02 \\
DDT24 & 1.26E-03 & 7.15E-02 & 7.06E-04 & 7.26E-03 & 2.10E-01 & 1.42E-01 & 3.54E-02 & 3.98E-02 & 2.20E-02 & 1.00E-02 & 8.00E-01 & 3.23E-02 \\
DDT40 & 5.33E-04 & 3.80E-02 & 2.62E-04 & 2.88E-03 & 1.35E-01 & 9.43E-02 & 2.38E-02 & 2.66E-02 & 1.59E-02 & 7.51E-03 & 9.69E-01 & 5.03E-02 \\
\noalign{\smallskip}
\tableline
\end{tabular}
}
\end{center}
\end{table*}

\placefigure{f1}
\begin{figure*}
\centering
\includegraphics[scale=0.45]{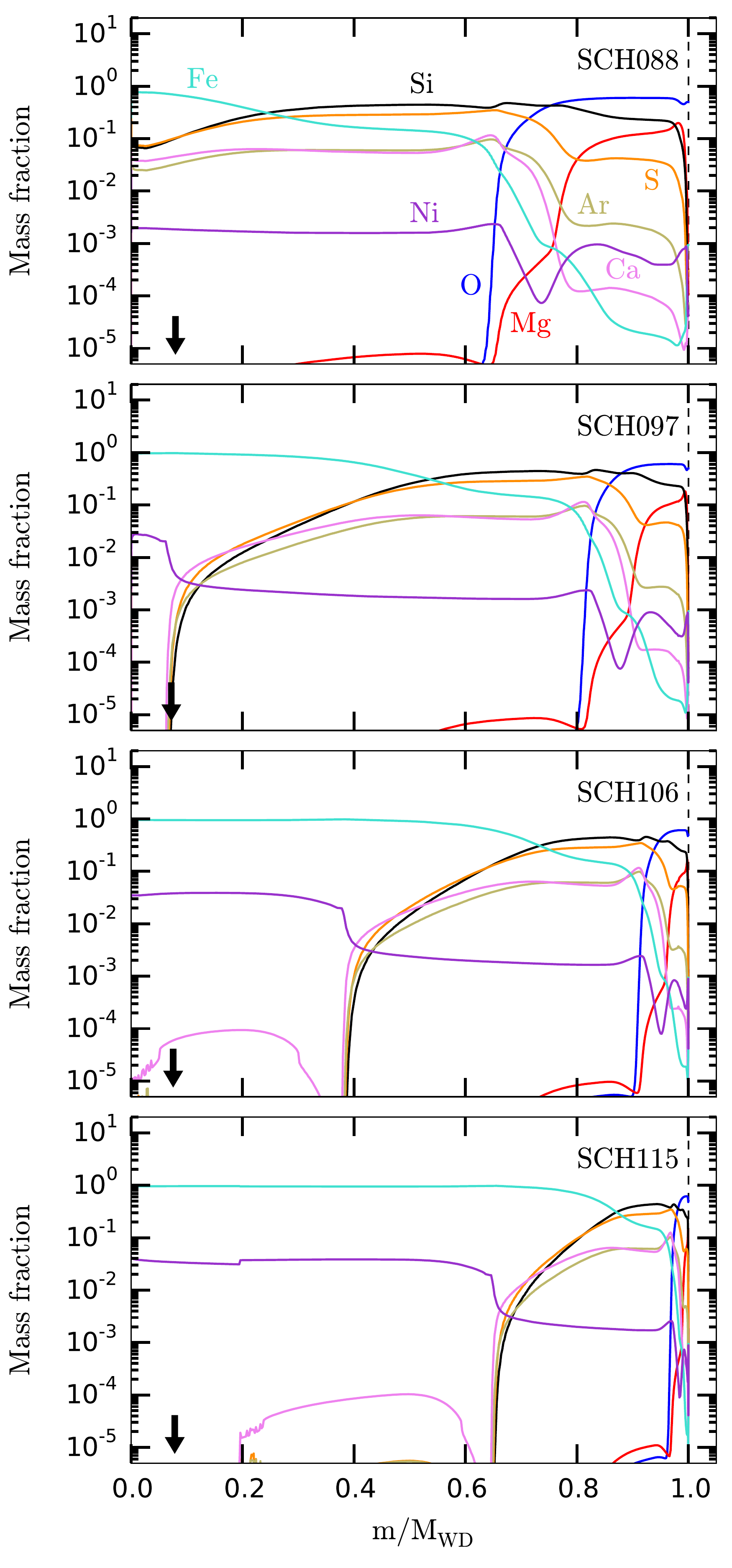}
\hspace{1.0 cm}
\includegraphics[scale=0.45]{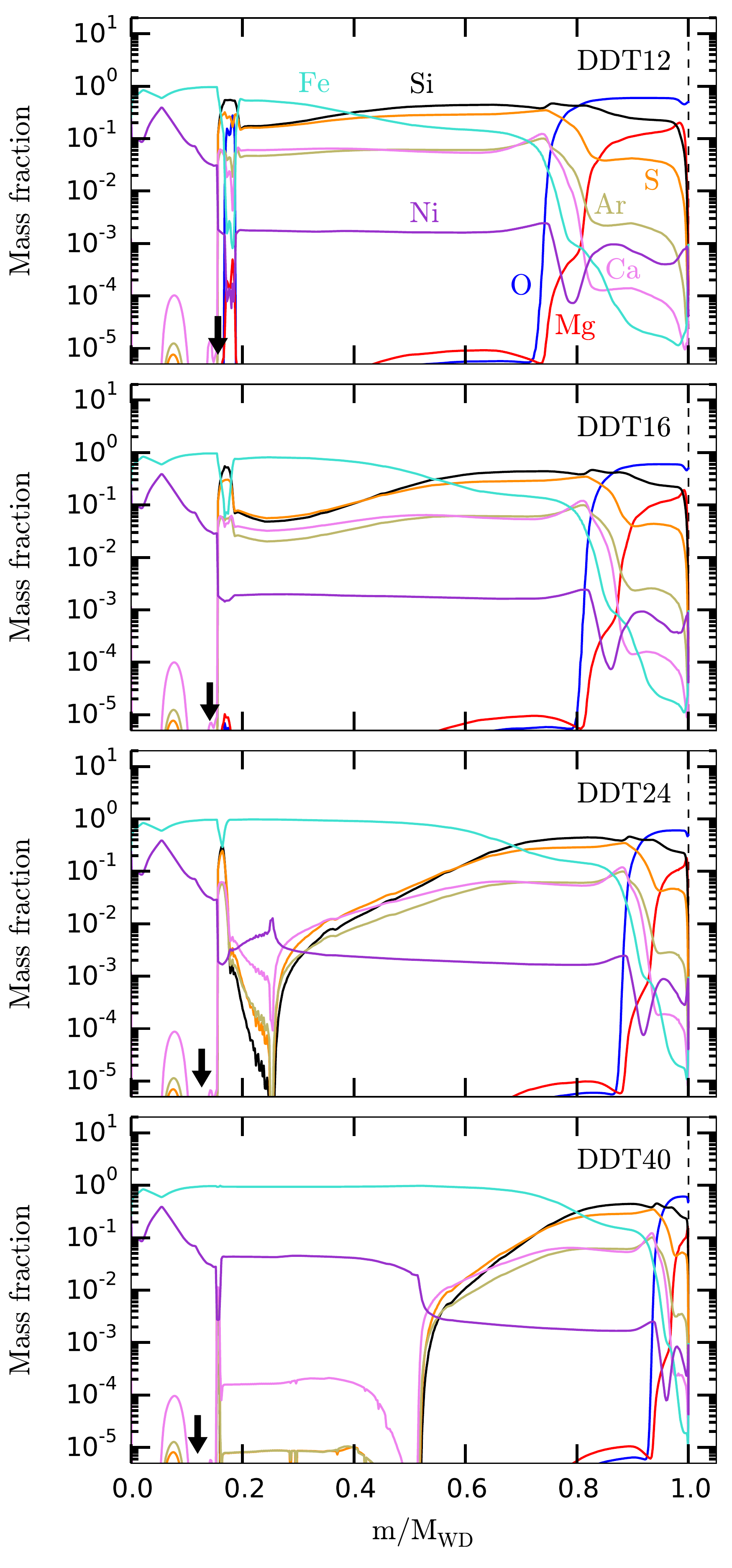}
\caption{Chemical composition for our SN Ia models listed in Table 
\ref{table:progenitor_models}. The vertical, dashed lines
indicate the outer surface of each ejecta model. The arrows depict the
locations of the RS at 538 years for 
$\rho_{\rm{amb}} \, =  \, 2 \times 10^{-24} \, \rm{g \, cm^{-3}}$ 
(see the discussion in Section \ref{subsec:syn_spectra}).}
\label{fig:Chemprof}
\end{figure*}

The X-ray emission from young (${\sim}$ a few 1000 years) SNRs is often-times 
dominated by strong emission lines from the shocked ejecta that can be used to 
probe the nucleosynthesis of the progenitor. These thermal 
(${\sim} \, 10^{7}\,$K) X-ray spectra are as 
diverse as their SN progenitors, and not even remnants of similar ages are 
alike. Their evolution and properties depend on various factors such as 
the structure and composition of the ejecta, the energy of the explosion, and 
the structure of the CSM that is left behind by the progenitor 
\citep[e.g.,][]{Ba03,Ba07,Pat12,Pat17,Woo17,Woo18}.

Therefore, young SNRs offer unique insights into both the supernova explosion 
and the structure of the ambient medium.
They are excellent laboratories to study the SN phenomenon 
\citep[e.g.,][]{Ba05,Ba06,Ba08b,Ba10,Vi12,Lee13,Lee14,Lee15,Sla14,Pat15}. 
The X-ray spectra of SNRs, unlike the optical spectra of SNe Ia, allow us to 
explore these issues without having to consider the complexities of radiative 
transfer \citep[e.g.,][]{Ste05,Ta11,Ash16,Wilk18}, because the plasma is at 
low enough density to be optically thin to its own radiation.

It is known that \mch\ models interacting with a uniform ambient medium can 
successfully reproduce the bulk properties of SNRs, such as ionization 
timescales \citep{Ba07}, Fe K$\alpha$ centroid energies, 
Fe K$\alpha$ luminosities \citep{Ya14a},
and radii \citep{PatB17}. However, there has been no exploration 
of the parameter space associated with the evolution of sub-\mch\ explosion 
models during the SNR stage for various dynamical ages. 
Here, we develop the first model grid of sub-\mch\ explosions in the SNR 
phase. We compare the bulk spectral and dynamical 
properties of \mch\ and sub-\mch\ models to the observed
characteristics of Galactic and Magellanic Cloud Ia SNRs.

This paper is organized as follows. In Section \ref{sec:method}, we describe
our hydrodynamical SNR models and the derivation of synthetic X-ray spectra. In 
Section \ref{sec:discussion}, we compare the bulk properties predicted by 
our model grid with observational data of Type Ia SNRs. Finally, in Section
\ref{sec:conclusions}, we summarize our results and outline future analyses 
derived from our work.

\section{Method}\label{sec:method}

\subsection{Supernova explosion models}\label{subsec:sn_models}

We use the spherically symmetric \mch\ and sub-\mch\ explosion models 
introduced in \citet{Ya15}, \citet{MR17} and \citet{McW18}, which are 
calculated with a version of the code described in \citet{Br12}, updated 
to account for an accurate coupling between hydrodynamics and nuclear 
reactions \citep{Br16, Br18}. 
The \mch\ models are delayed detonations 
\citep{Kh91} with a central density 
$\rho_{\rm{c}} = 3 \times 10^{9} \, \rm{g \, cm^{-3} }$, 
deflagration-to-detonation densities 
$\rho_{\rm{DDT}}$ $\rm{[10^{7} \, g \, cm^{-3}]} =$
 $1.2, 1.6, 2.4, 4.0$ and kinetic energies 
 $ E_{k} \, [10^{51} \, \rm{erg}] = 1.18, 1.31, 1.43, 1.49$. 
 They are similar to the models DDTe, DDTd, DDTb, and DDTa 
 ($\rho_{\rm{DDT}}$ $\rm{[10^{7} \, g \, cm^{-3}]} =$
$1.3, 1.5, 2.6, 3.9$) by \citet{Ba03,Ba05,Ba06,Ba08b}. We label these 
explosions as DDT12, DDT16, DDT24, and DDT40.

The sub-\mch\ models are central 
 detonations of C/O WDs with a core temperature 
$T_{\rm{c}} \, [\rm{K}] = 10^{8}$, 
masses $ M_{\rm{WD}} \, [M_{\odot}] = 0.88, 0.97, 1.06, 1.15$,
and kinetic energies $ E_{k} \, [10^{51} \, \rm{erg}] = 0.92, 1.15, 1.33, 1.46$, 
similar to the models by \citet{Si10}. We label these explosions as 
SCH088, SCH097, SCH106, and SCH115. For both sets of models, the progenitor 
metallicity is $Z = 0.009$ ($0.64 \, Z_{\odot}$ taking 
$Z_{\odot} = 0.014$, \citealt{As09}). 
We choose this value because it is close to 
the metallicity $Z = 0.01$ employed by \citet{Ba03,Ba05,Ba06,Ba08b} in 
their \mch\ progenitors. The intermediate-mass elements (Si, S, Ar, Ca) 
are produced in the outer region of the exploding WDs, whereas the 
iron-peak elements (Cr, Mn, Fe, Ni) are synthesized in the inner layers. 
Table \ref{table:progenitor_models} presents the total yields for some 
representative elements in these \mch\ and sub-\mch\ models. Figure
\ref{fig:Chemprof} shows the chemical profiles as a function of the
enclosed mass for each model.

\subsection{Supernova remnant models}\label{subsec:snr_models}

\placefigure{f2}
\begin{figure}
\centering
\hspace{-1.5 cm}
\includegraphics[scale=0.65]{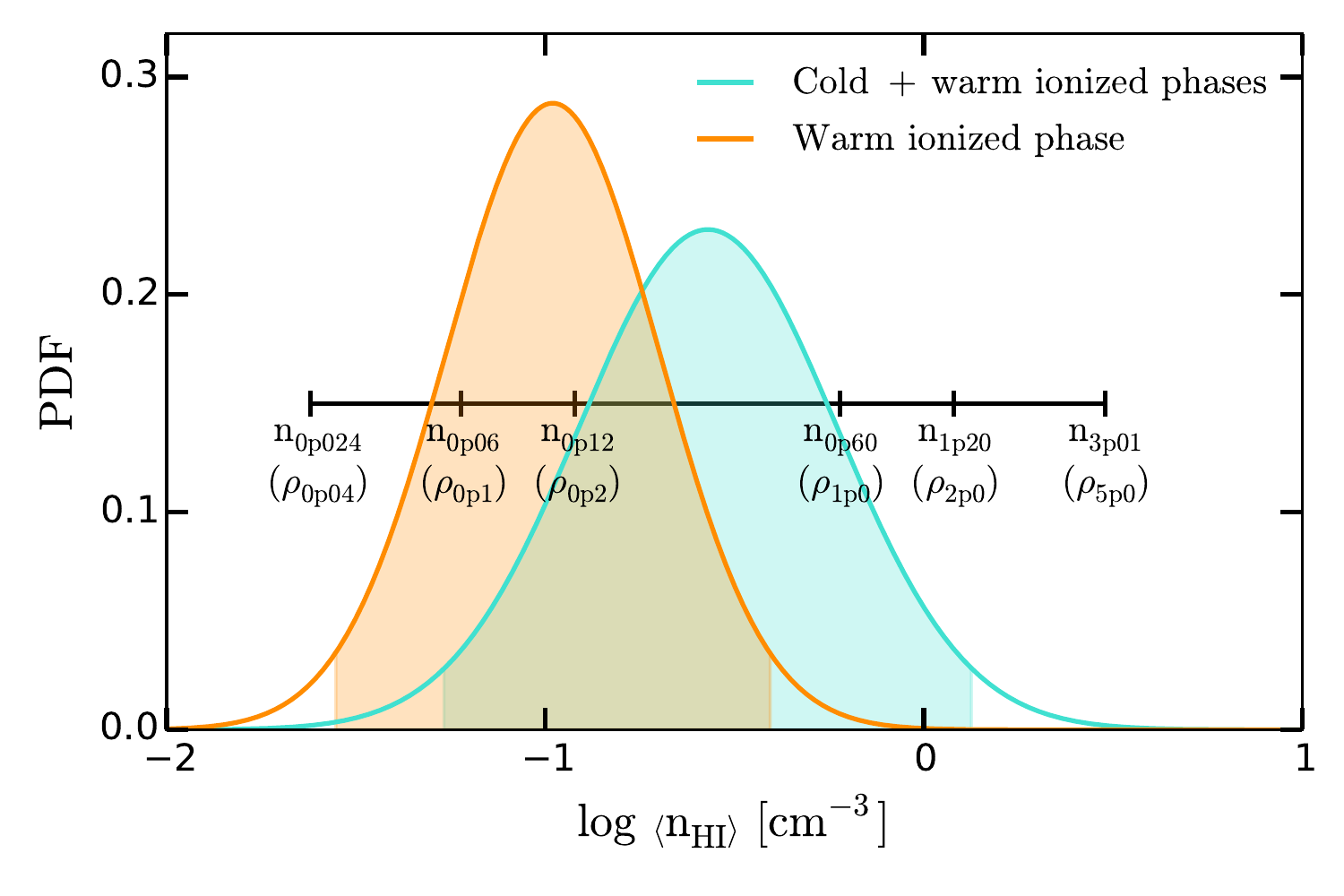}
\caption{Log-normal probability distribution functions (PDFs) for 
the diffuse gas in the Milky Way \citep{Ber08}. The shaded contours
represent the 2$\sigma$ regions for each PDF. The six $n_{\rm{amb}}$
values used in this work ($0.024, 0.06, 0.12, 0.60, 
1.20, 3.01 \, \rm{cm^{-3}}$) 
are depicted along a black, horizontal line.}
\label{fig:ISM}
\end{figure}

We study the time evolution of these SN Ia models with a self-consistent 
treatment of the nonequilibrium ionization (NEI) conditions in young SNRs 
performed by the cosmic ray-hydro-NEI code, hereafter \texttt{ChN} 
\citep{Ell07,Pat09,Ell10,Pat10,Cas12,Lee12,Lee14,Lee15}. 
\crcode\ is a one-dimensional Lagrangian hydrodynamics code based on the multidimensional 
code \texttt{VH-1} \citep[e.g.,][]{Blo93}. \crcode\ simultaneously calculates 
the thermal and nonthermal emission at the FS and RS in the expanding SNR 
models. It couples hydrodynamics, NEI calculations, plasma emissivities, 
time-dependent photoionization, radiative cooling, forbidden-line emission, 
and diffusive shock acceleration, though we do not include diffusive shock 
acceleration in our calculations. \crcode\ is a tested, flexible code that 
has successfully been used to model SNRs in several settings 
\citep[e.g.][]{Sla14,Pat15}.

Young Ia SNRs are in NEI because, at the low 
densities involved ($n \, {\sim} \, 1 \, \rm{cm}^{-3}$), not enough time 
has elapsed since the ejecta were shocked to equilibrate the ionization and 
recombination rates \citep{Itoh77,Ba10}. Consequently, these NEI plasmas are 
underionized when compared to collisional ionization equilibrium plasmas
\citep{Vi12}. 
The shock formation and initial plasma heating do not stem from Coulomb interactions, 
but from fluctuating electric and magnetic fields in these so-called collisionless 
shocks \citep[][]{Vi12}. In the ISM, the mean free path and the typical ages
for particle-to-particle interactions are larger than those of SNRs 
($\approx 10^{2}-10^{3} \, \rm{years}, \approx 1-10 \, \rm{pc}$).

The efficiency of electron heating at the shock transition, 
i.e., the value of $\beta = T_{e} / T_{i}$ at the shock, is not well determined 
\citep[see, e.g.,][]{Bo01}. In principle, the value of $\beta$ can range between 
$\beta = \beta_{\rm{min}} = m_{e} / m_{i}$ and full equilibration ($\beta = 1$), 
with partial equilibration being the most likely situation 
\citep[$\beta_{\rm{min}} < \beta < 1$,][]{Bo01,Gha07,Ya14b}. Here we set 
$\beta = \beta_{\rm{min}}$ for illustration purposes, even though previous studies 
\citep[e.g.,][]{Ba05,Ba06,Ya14a} have shown that $\beta$ has an important effect on 
the Fe K$\alpha$ luminosities. This can be critical when trying to fit an SNR spectrum 
with a specific model, but here we are just interested in the bulk properties of the 
models, and we defer detailed fits to future work.

We consider uniform ambient media composed of hydrogen 
\citep[$\rho_{\rm{amb}} = m_{H} \, n_{\rm{amb}}$, e.g.][]{Ba03,Ba06,Ba08b,PatB17} 
with a range of densities: 
$\rho_{\rm{amb}} \, [10^{-24} \, \rm{g \, cm^{-3}}] = 0.04, 0.1, 0.2, 1.0, 2.0, 5.0 
\equiv $ $n_{\rm{amb}} \, [\rm{cm^{-3}}] = 0.024,0.06, 0.12, 0.60, 1.20, 3.01$. 
We label each SNR model from the SN model and ambient medium density, e.g. 
SCH115\_0p04, SCH115\_0p1, SCH115\_0p2, 
SCH115\_1p0, SCH115\_2p0, and SCH115\_5p0. We have chosen these ambient 
densities to be in the same range considered by \citet{Pat15}. The three 
highest densities were used in the studies by \citet{Pat12} and 
\citet{Ya14a}, so we will be able to compare our results to theirs. This 
makes a total of 48 SNR models that we evolve up to an expansion age of 
5000 years. For each SNR model, we record a total of 30 time epochs, 
starting at 105 years. The time bins are linearly spaced at young ages 
and smoothly become logarithmically spaced at late ages. We also 
record 30 Lagrangian profiles in linearly spaced time bins for each model.

Our choice of ambient medium densities is motivated by observations of the 
ISM in the Milky Way. Interstellar gas can be found in five different phases 
\citep{Fe98,Fe01}: molecular 
($T_{mol} \, {\sim} \ 10-20 \ \rm{K}$,  $n_{mol} \ {\sim} \ 10^{2}-10^{6} \ \rm{cm}^{-3}$),
cold neutral 
($T_{cold} \, {\sim} \ 50-100 \ \rm{K}$,  $n_{cold} \ {\sim} \ 20-50 \ \rm{cm}^{-3}$), 
warm neutral 
($T_{warm,n} \, {\sim} \ 6000-10000 \ \rm{K}$,  $n_{warm,n} \ {\sim} \ 0.2-0.5 \ \rm{cm}^{-3}$), 
warm ionized 
($T_{warm,i} \, {\sim} \ 8000 \ \rm{K}$,  $n_{warm,i} \ {\sim} \ 0.2-0.5 \ \rm{cm}^{-3}$), and 
hot ionized 
($T_{hot} \, {\sim} \ 10^{6} \ \rm{K}$,  $n_{hot} \ {\sim} \ 0.0065 \ \rm{cm}^{-3}$). 
Among these, the warm ionized phase  
has the highest filling factor and therefore is the most likely environment for 
Type Ia SNRs. \citet{Wol03} gives a mean value for the neutral hydrogen density 
in the Galactic disk $\left<n_{H I}\right> \, = 0.57 \ \rm{cm}^{-3}$. More 
recently, \citet{Ber08} fit log-normal distributions to the diffuse gas in the 
MW centered on $\left<n_{H I}\right> \, \approx 0.3 \ \rm{cm}^{-3}$ (cold and 
warm ionized) and $\left<n_{H I}\right> \, \approx 0.1 \ \rm{cm}^{-3}$ (warm 
ionized). We compare these distributions to our uniform density values 
in Figure \ref{fig:ISM}.

\placefigure{f3}
\begin{figure}
\centering
\hspace{-1.5 cm}
\includegraphics[scale=0.37]{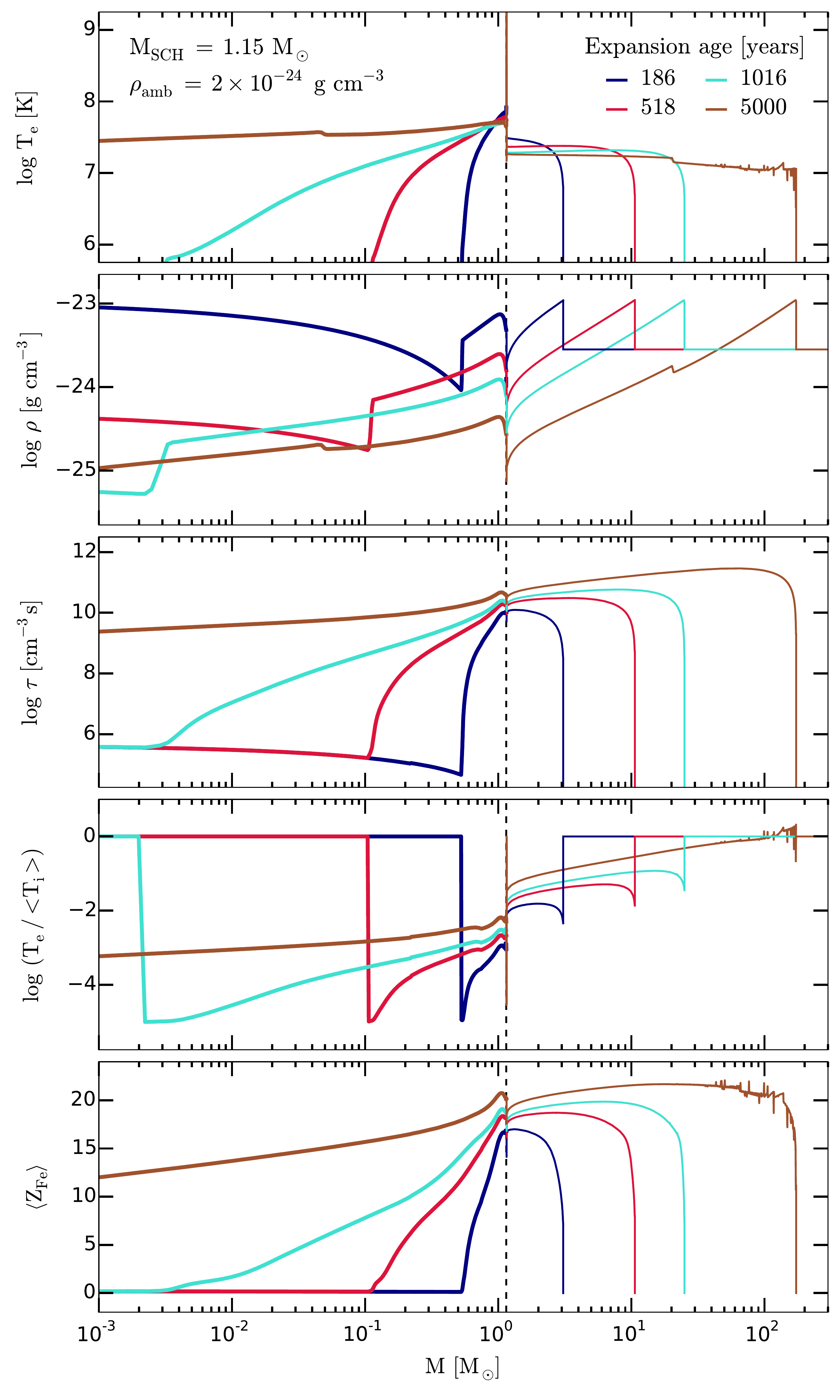}
\caption{Time evolution of the electron temperature $T_e$, density $\rho$,
ionization timescale $\tau = n_{e} t$, average efficiency of
post-shock equilibration $T_{e} / \left<T_{i}\right>$ 
and average iron effective charge state 
$\left<z_{\rm{Fe}}\right>$  profiles
as a function of the enclosed mass for model SCH115\_2p0.
The CD between the ejecta (thick lines)
and the ambient medium swept up by the FS (thin lines)
is depicted as a dashed, black vertical line, located at $1.15 \, M_{\odot}$. 
The spatial location of the RS can be 
appreciated in the navy ($\sim 0.55 \, M_{\odot}$), the crimson
($\sim 0.1 \, M_{\odot}$) and the turquoise 
($\sim 0.002 \, M_{\odot}$) profiles.}
\label{fig:Hydro}
\end{figure}

Figure \ref{fig:Hydro} shows the profile time evolution for 
 a fiducial model, explosion progenitor SCH115 with an ambient density 
$\rho_{\rm{amb}} \, =  \, 2 \times 10^{-24} \, \rm{g \, cm^{-3}}$.
The profiles for 186 (navy), 518 (crimson), and 1016 (turquoise)
years show the RS propagation toward the center of the SNR. After reaching 
the center, the RS bounces back and moves outwards into the previously 
shocked ejecta, creating more reflected shocks when it reaches the contact 
discontinuity (CD). This effect can be seen in the first and the second 
panel of Figure \ref{fig:Hydro} ($T_{e}$ versus $M$, $\rho$ versus $M$) 
around $M \, {\sim} \, 0.05 \, M_{\odot}$ and 
$M \, {\sim} \, 20 \, M_{\odot}$ at 5000 years (brown).

$T_e$ increases with time in the inner layers after they are swept by 
the RS. As the SNR expands, the density $\rho$
of the shocked ejecta and ISM decreases steadily, and therefore the 
electron density $n_{e}$ diminishes with time. 
In \crcode, the unshocked plasma is assumed to be 10\% singly ionized.

The salient features in the evolution of this particular SNR model are 
representative of the entire grid. The ejecta with the highest ionization 
state are always found close to the contact discontinuity (CD), since they were 
shocked at an earlier age and higher density. Because this is also the 
densest region at all times, it has the highest emission measure and 
thus will dominate the spatially integrated X-ray emission. However, since the 
chemical composition of SN Ia ejecta is markedly stratified, it is often 
the case that different chemical elements sample different parts of the 
SNR structure, and therefore show different ionization timescales and 
electron temperatures \citep[see the discussions in][]{Ba03,Ba05}. This 
feature of the models is in good agreement with observations of young 
SNRs \citep[e.g.][]{Ba07}.

\placefigure{f4}
\begin{figure*}
\centering
\includegraphics[scale=0.70]{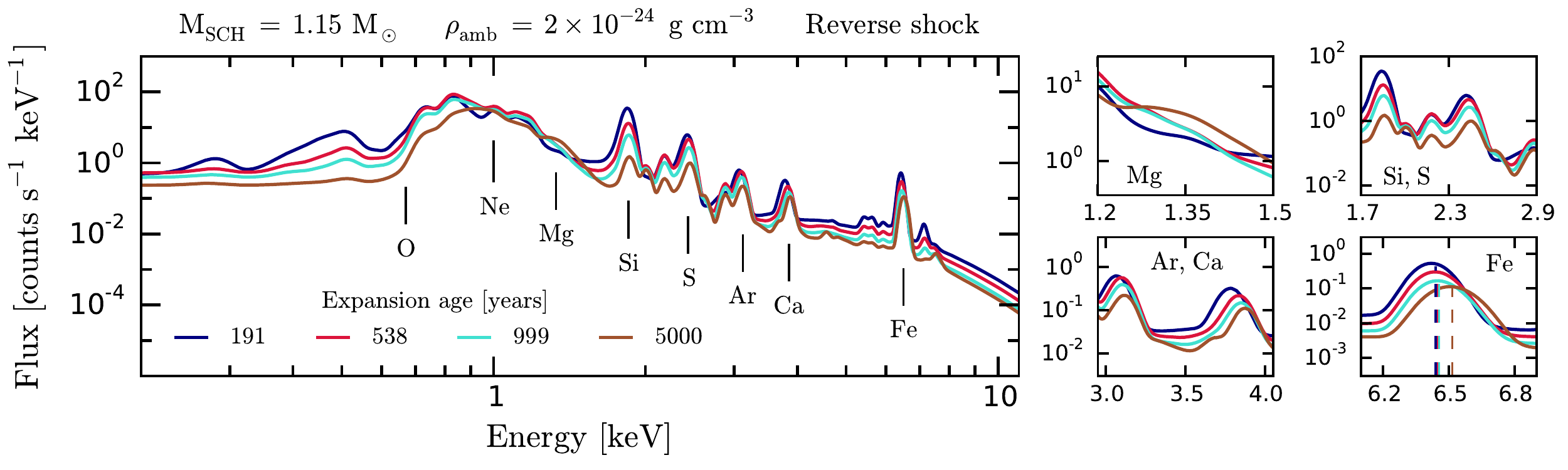}
\caption{Integrated RS synthetic spectra normalized to $D = 10$ kpc for the model shown in Figure 
\ref{fig:Hydro} at the nearest time snapshots (see the explanation in the text). 
The relevant atomic transitions are labeled. The zoomed boxes depict different energy regions:
Mg (up left), Si, S (up right), Ar, Ca (low left), and Fe (low right). The latter shows
the time evolution of the Fe K$\alpha$ centroid energy (dashed, vertical lines).}
\label{fig:Spectra_ages}
\end{figure*}

\placefigure{f5}
\begin{figure*}
\centering
\includegraphics[scale=0.70]{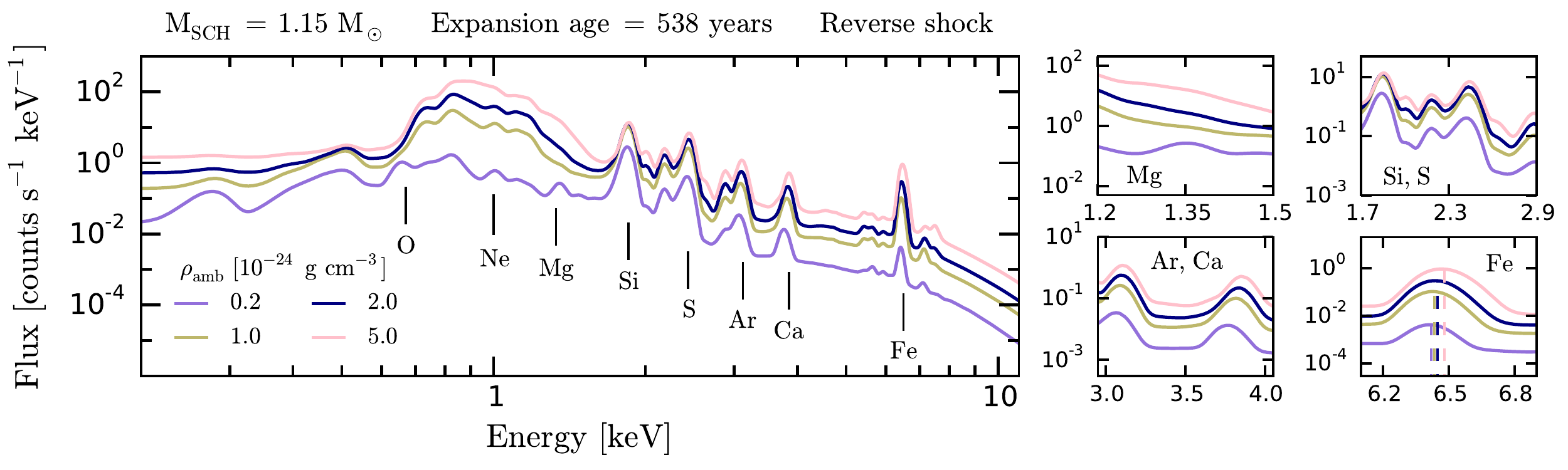}
\caption{Integrated RS synthetic spectra normalized to $D = 10$ kpc for model SCH115, 
for the four highest ambient densities ($\rho_{0p2}$, $\rho_{1p0}$, $\rho_{2p0}$, $\rho_{5p0}$)
 and a fixed expansion age of 538 years. 
The zoomed boxes are identical to those of Figure \ref{fig:Spectra_ages}.}
\label{fig:Spectra_densities}
\end{figure*}

\placefigure{f6}
\begin{figure*}
\centering
\includegraphics[scale=0.70]{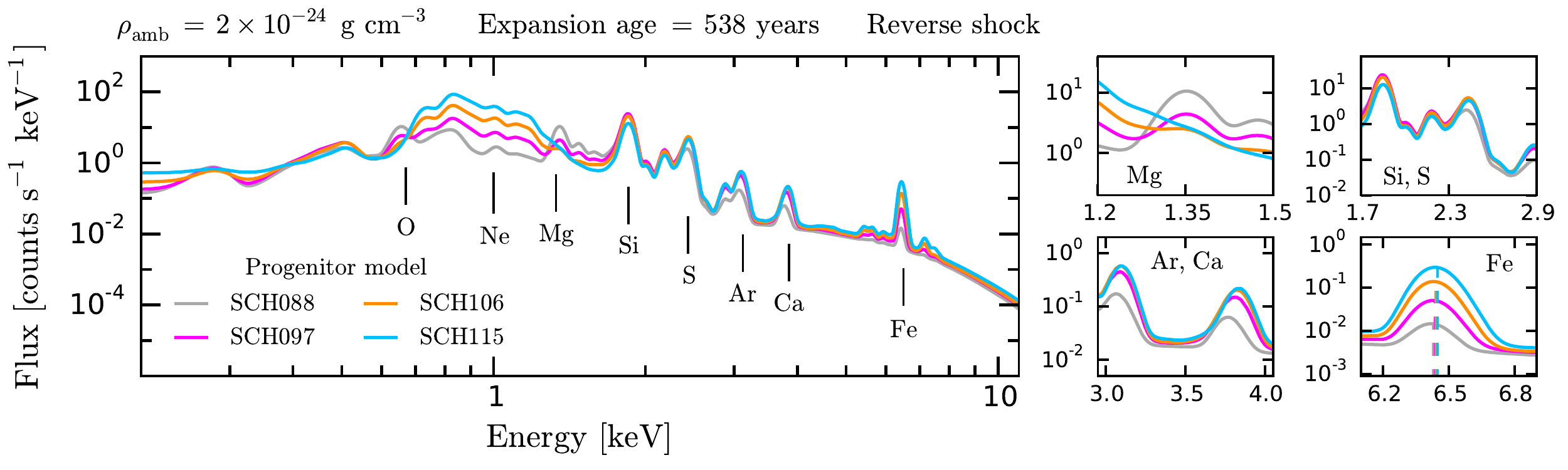}
\caption{Integrated RS synthetic spectra normalized to $D = 10$ kpc for models SCH088, 
SCH097, SCH106, and SCH115 at a fixed expansion age of 538 years and a fixed ambient 
density $\rho_{\rm{amb}} \, =  \, 2 \times 10^{-24} \, \rm{g \, cm^{-3}}$. 
The zoomed boxes are identical to those of Figure \ref{fig:Spectra_ages}.}
\label{fig:Spectra_models_SCH}
\end{figure*}

\placefigure{f7}
\begin{figure*}
\centering
\includegraphics[scale=0.70]{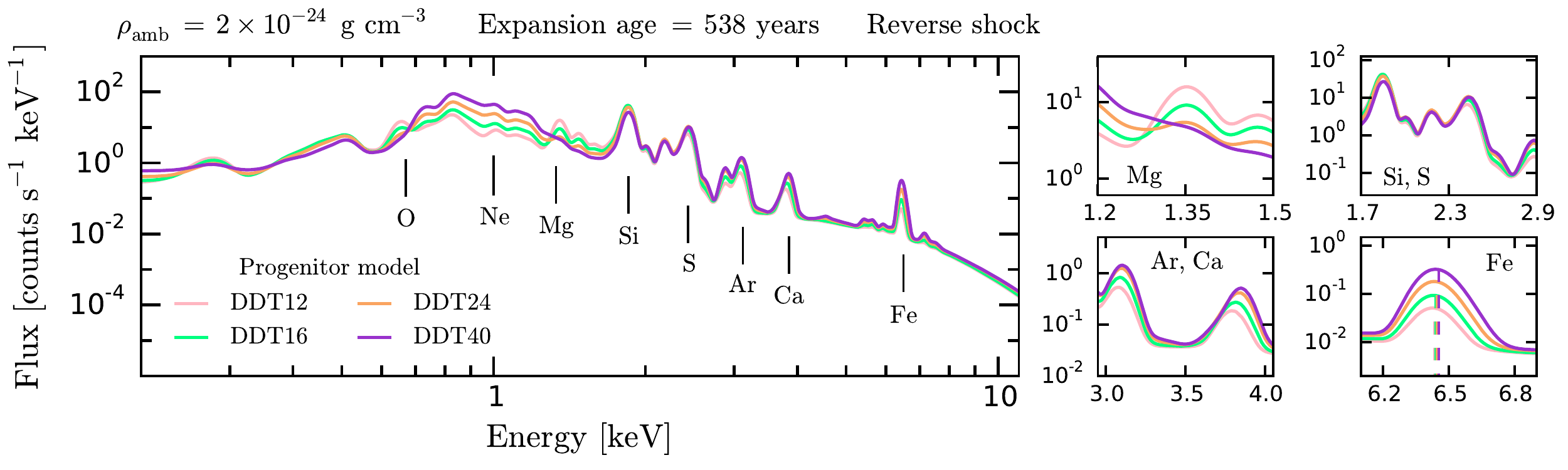}
\caption{Integrated RS synthetic spectra normalized to $D = 10$ kpc for models DDT12, 
DDT16, DDT24, and DDT40 at a fixed expansion age of 538 years and a fixed ambient 
density $\rho_{\rm{amb}} \, =  \, 2 \times 10^{-24} \, \rm{g \, cm^{-3}}$. 
The zoomed boxes are identical to those of Figure \ref{fig:Spectra_ages}.}
\label{fig:Spectra_models_DDT}
\end{figure*}

\subsection{Synthetic spectra}\label{subsec:syn_spectra}

Our ejecta models determine the masses, chemical abundances, and initial 
velocities for each mass layer. We consider 19 elements: H, He, C, N, O, 
Ne, Na, Mg, Al, Si, P, S, Ar, Ca, Ti, Cr, Mn, Fe, and Ni, with a total of 
297 ions. For each ion species $I$ corresponding to an element $X$,
we calculate the differential emission measure (DEM) in 51 equally log-spaced 
$T_{e}$ bins between $10^{4}$ and $10^{9}$ K, normalized 
to a distance of $D = 10$ kpc \citep{Ba03,Ba06}:

\begin{equation}
(\rm{DEM})_{I,X} = n_{I} \, n_{e} \times \dfrac{dV}{dT_{e}} 
\times \dfrac{1}{4 \pi D [cm]^{2}} \times \dfrac{10^{-14}}{\rm{angr}(X)} 
\end{equation}

where $n_{I}, n_{e}$ are the ion and electron densities, $dV$ is the 
volume element for each layer, $\rm{angr}(X)$ are the \texttt{XSPEC} \citep{Ar96} 
default conversion factors for the solar abundances \citep{AG89} and $10^{-14}$ is 
a normalization applied to the emissivities in \texttt{XSPEC}. We couple 
these DEMs to the atomic emissivity code \texttt{PyAtomDB} 
\citep[\texttt{AtomDB} version 3.0.9; see, e.g.,][]{Fo12,Fo14} in order to 
calculate the emitted flux for each model at a given photon energy. We separate 
the RS and the FS contribution and generate nonconvolved photon spectra in 
10000 equally spaced bins of size 1.2 eV between 0.095 and 12.094 keV.
Thermal broadening and line splitting due to bulk motions are ignored in this 
version of the synthetic spectra, but we plan to include them in future versions.

\begin{table*}
\begin{center}
\caption{Data corresponding to the Ia SNRs in our sample. \label{table:remmants}}
\begin{tabular}{cccccccc}
\tableline
\noalign{\smallskip}
\tableline
\noalign{\smallskip}
Name
& $E_{\, \rm{Fe}_{K\alpha}}$\footnote{Centroid energies and fluxes from \cite{Ya14a}, except for G1.9+0.3 \citep{Bo13} and DEM L71 \citep{Mag16}, who report luminosities.\label{foot:Yam}}
& $F_{\, \rm{Fe}_{K\alpha}}$\footref{foot:Yam} & Distance & $L_{\, \rm{Fe}_{K\alpha}}$ & Radius\footnote{For remnants with distance uncertainties, we calculate their radii using the angular diameters listed in Table 1 from \cite{Ya14a}.} & Age & References\footnote{Representative references: 
(1) \cite{Reyn99}; (2) \cite{San05}; (3) \cite{Rey07}; (4) \cite{Park13}; (5) \cite{Sa05}; (6) \cite{Lea16}; (7) \cite{Ba06}; (8) \cite{Tia11}; (9) \cite{Wi11a}; (10) \cite{Ya12a}; (11) \cite{Cas13};
(12) \cite{Ya08}; (13) \cite{Ra06}; (14) \cite{Ya12b}; (15) \cite{Gia09}; (16) \cite{Pan14}; (17) \cite{Le03}; (18) \cite{Res05}; (19) \cite{Wi14};  (20) \cite{War04};  (21) \cite{Res08}; (22) \cite{Kos10}; (23) \cite{Rey08}; (24) \cite{Bo13}; (25) \cite{Hu03}; (26) \cite{vaH03}; (27) \cite{Mag16}.} \\
\noalign{\smallskip}
& $\mathrm{eV}$ & ($\mathrm{10^{-5}\,ph\,cm^{-2}\,s^{-1}}$) & ($\mathrm{kpc}$)  & ($\mathrm{10^{40}\,ph\,s^{-1}}$) & ($\mathrm{pc}$) & ($\mathrm{years}$) &  \\
\noalign{\smallskip}
\tableline
\noalign{\smallskip}
			
Kepler & $6438 \pm 1$ & $34.6 \pm 0.2$ & $3.0-6.4$ & $91 \pm 66$ & $2.3 \pm 0.9$ & $414$ & 1, 2, 3, 4  \\
			
3C 397 & $6556 ^{+ 4} _{-3}$ & $13.7 \pm 0.4$ & $6.5-9.5$ & $105 \pm 39$ & $5.3 \pm 0.5$ & $1350-5300$ & 5, 6  \\
			
Tycho & $6431 \pm 1$ & $61.0 \pm 0.4$ & $2.5-3.0$ & $55 \pm 10$ & $3.3 \pm 0.3$ & $446$ & 7, 8  \\
			
RCW 86 & $6408 ^{+ 4} _{-5}$ & $14.0 \pm 0.7$ & $2.5$ & $10.5 \pm 0.5$ & $16$ & $1833$ & 9, 10, 11  \\
			
SN 1006 & $6429 \pm 10$ & $2.55 \pm 0.43$ & $2.2$ & $1.5 \pm 0.3$ & $10$ & $1012$ & 12  \\
			
G337.2$-$0.7 & $6505 ^{+ 26} _{-31}$ & $0.21 \pm 0.06$ & $2.0-9.3$ & $0.8 \pm 1.1$ & $4.9 \pm 3.2$ & $5000-7000$ & 13  \\
			
G344.7$-$0.1 & $6463 ^{+ 9} _{-10}$ & $4.03 \pm 0.33$ & $14$ & $95 \pm 8$ & $16$ & $3000-6000$ & 14  \\
			
G352.7$-$0.1 & $6443 ^{+ 8} _{-12}$ & $0.82 \pm 0.08$ & $7.5$ & $5.5 \pm 0.5$ & $6$ & ${\sim} \, 1600$ & 15, 16   \\
			
N103B & $6545 \pm 6$ & $2.15 \pm 0.10$ & 50\footnote{Distance to the Large Magellanic Cloud (LMC) from \cite{Piet13}. 
\label{foot:LMC} \\} & $643 \pm 30$ & $3.6$ & ${\sim} \, 860$ & 17, 18, 19  \\
			
0509$-$67.5 & $6425 ^{+ 14} _{-15}$ & $0.32 \pm 0.04$ & 50\footref{foot:LMC} & $96 \pm 12$ & $3.6$ & ${\sim} \, 400$ & 18, 20, 21  \\
			
0519$-$69.0 & $6498 ^{+ 6} _{-8}$ & $0.93 \pm 0.05$  & 50\footref{foot:LMC} & $278 \pm 15$ & $4.0$ & ${\sim} \, 600$ & 18, 21, 22  \\
			
G1.9+0.3 & $6444$ & - & ${\sim} \, 8.5$ & 1 & ${\sim} \, 2.0$ & ${\sim} 150$ &  23, 24 \\
			
DEM L71 & $6494 \pm 58$ & - & 50\footref{foot:LMC} & $26 ^{+ 8} _{-9}$ & 8.6 & ${\sim} \, 4700$ & 25, 26, 27  \\
			
\noalign{\smallskip}
\tableline
\end{tabular}
\end{center}
\end{table*}

We generate synthetic spectra for both RS and FS convolved with the \textit{Suzaku} 
spectral and ancillary responses \citep{Mit07}. We choose 
\textit{Suzaku} over \textit{Chandra} or \textit{XMM--Newton} for illustration 
purposes, given its superior spectral resolution around the K$\alpha$ transitions 
from Fe-peak elements ($\approx 5.5-8.0$ keV). For simplicity, we do not include the 
effect of interstellar absorption (relevant below $\sim$ 1 keV). In any case, most Ia 
SNRs have column densities smaller than $10^{22} \, \rm{cm^{-2}}$ 
\citep[e.g.,][]{Le03,War04,Ba06,Rey07,Kos10,Ya14b}. All the convolved and nonconvolved 
spectra are publicly available in a repository (\url{https://github.com/hector-mr}).

\placefigure{f8}
\begin{figure*}
\centering
\includegraphics[scale=0.37]{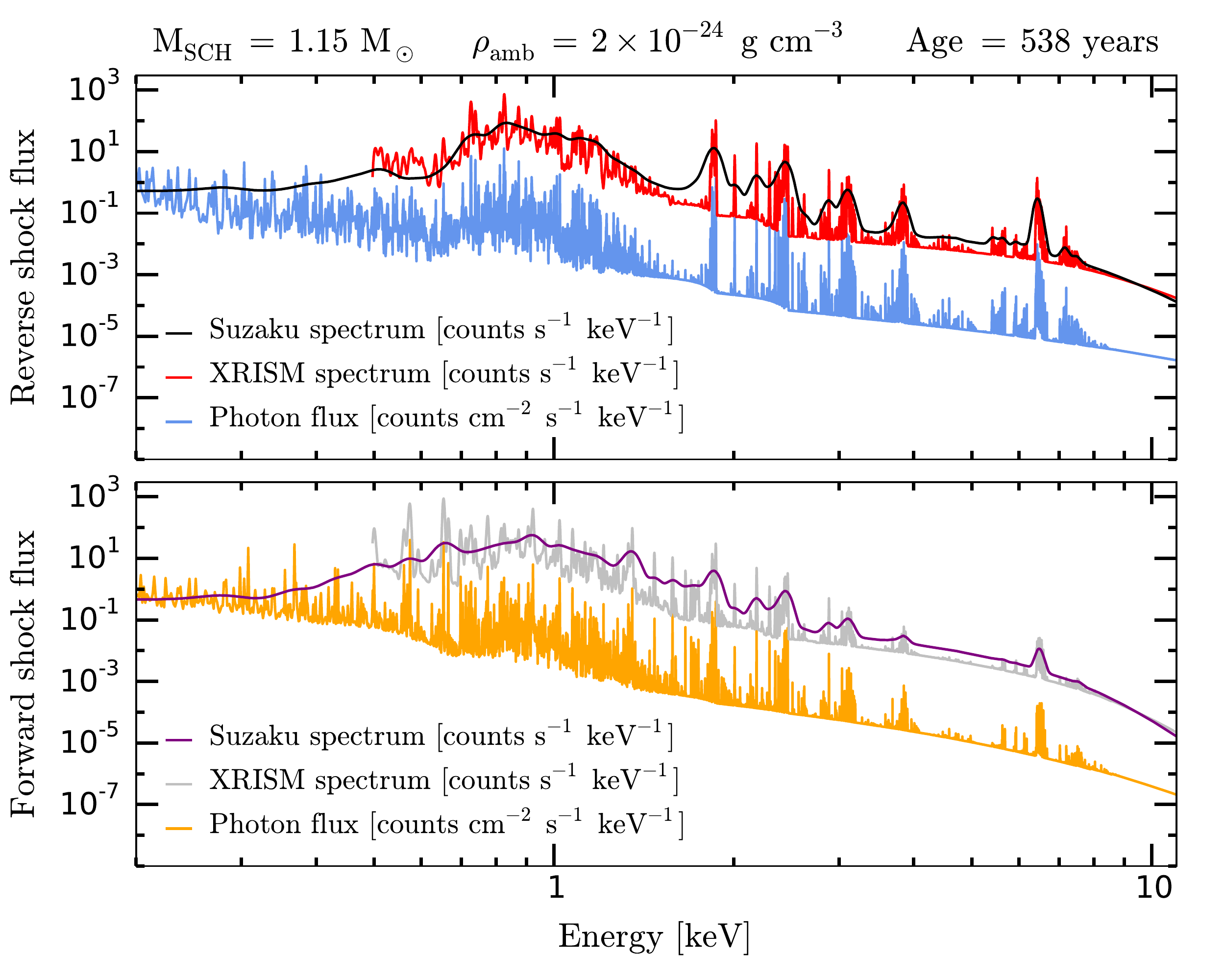}
\includegraphics[scale=0.268]{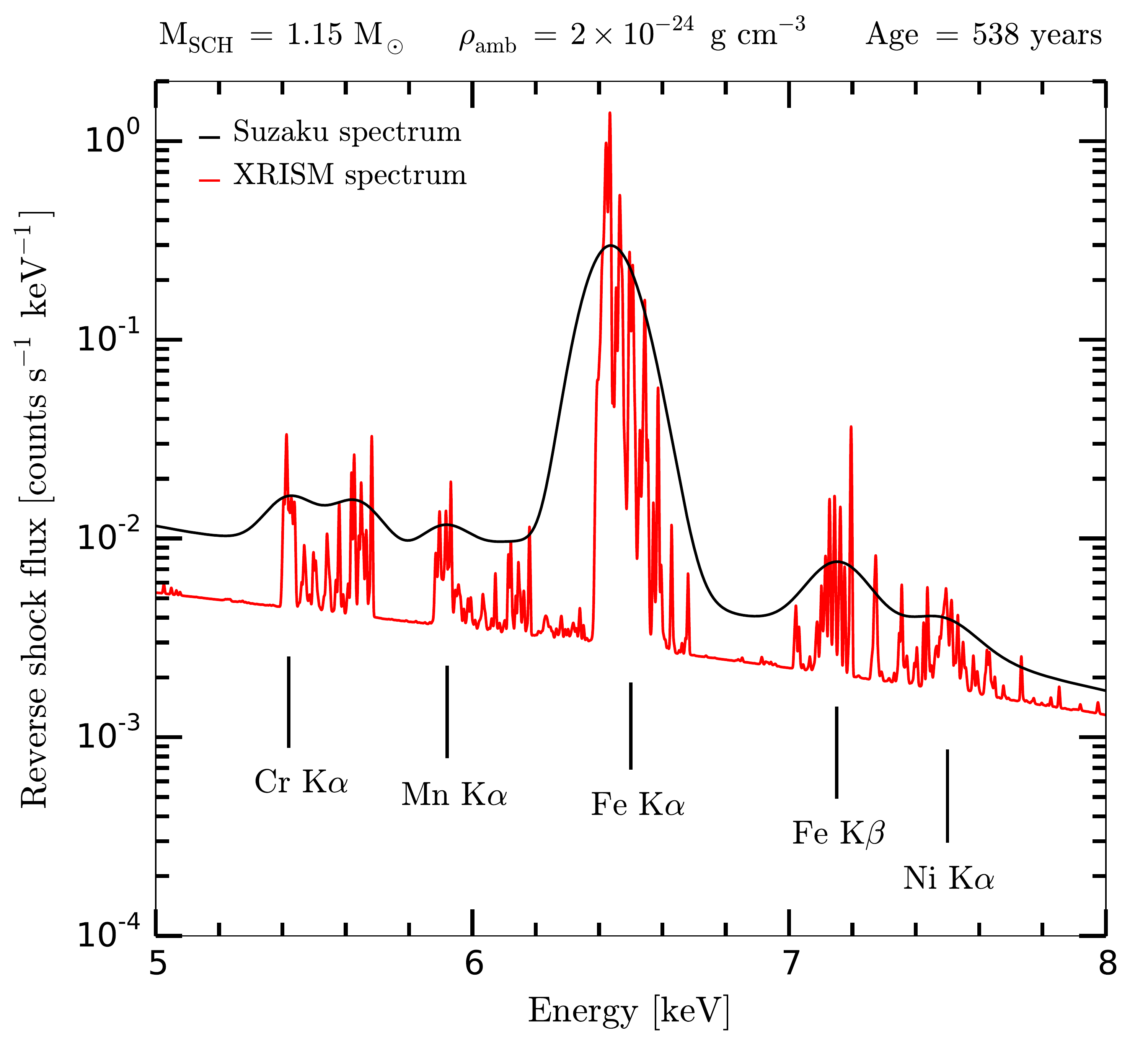}
\caption{Left: Photon, \textit{Suzaku} and \textit{XRISM} spectra for model 
SCH115\_2p0 at a fixed expansion age of 538 years (Top: Reverse shock. Bottom: 
Forward shock). Right: Zoomed-in reverse shock spectra around the Fe K$\alpha$ 
complex. The relevant atomic transitions are labeled.}
\label{fig:Sp_phSuXA}
\end{figure*}

Figure \ref{fig:Spectra_ages}  shows the time evolution of the X-ray flux from the RS 
for the fiducial model shown in Figure \ref{fig:Hydro}. We do not show the thermal 
spectrum from the FS because it is very weak or absent in many young Type Ia SNRs, 
often being replaced by nonthermal synchrotron emission \citep[e.g.,][]{War04,War05,Ca08}. 
While the \crcode\ code has the capability to model the modification of the FS dynamics 
and spectrum due to particle acceleration processes \citep[e.g.,][]{Sla14}, 
this falls outside the scope of the present work. The thermal RS flux shown in Figure
\ref{fig:Spectra_ages} decreases with time because the ejecta density decreases steadily, 
and the emission measure scales as $n_{e}^{\, 2}$. This effect usually dominates over the 
steady increase in $T_{e}$ due to electron-ion collisions in the shocked plasma 
(see Figure \ref{fig:Hydro}), which tends to increase the emitted flux. The centroids of 
the K$\alpha$ transitions move to higher energies with time, especially for Ca, Fe, and Ni, 
because those elements have a large range of charge states. For elements with lower 
atomic numbers, like Si and S, the centroid energies saturate when the He-like ions become 
dominant, and then the Ly$\alpha$ transitions from H-like ions begin to appear. For 
this fiducial model, the spectrum at 5000 years (brown) shows a Ti K$\alpha$ feature at 
$\approx 4.5$ keV.

Figure \ref{fig:Spectra_densities} shows the effect of varying the ambient medium density 
on the RS spectra for the same explosion model (SCH115) at a fixed expansion age of 538 
years. Higher $\rho_{\rm{amb}}$ translate into higher ejecta densities due to a slower
ejecta expansion. This yields higher fluxes and centroid energies for all transitions due 
to the increased rate of ionizing collisions. As $\rho_{\rm{amb}}$ increases, the Fe L-shell 
transitions dominate the flux around ${\sim} \,$1 keV. Figures \ref{fig:Spectra_models_SCH} and 
\ref{fig:Spectra_models_DDT} show the RS spectra for all sub-\mch\ and \mch\ progenitor 
models with the same $\rho_{\rm{amb}} \, \left(2 \times 10^{-24} \, \rm{g \, cm^{-3}}\right)$ 
and expansion age (538 years). The differences between the models are largest in the bands 
dominated by the Fe L-shell and K-shell transitions. This is due to the different 
distribution of Fe-peak elements in the inner ejecta region for different models. In 
sub-\mch\ models with larger masses and \mch\ models with higher DDT transition densities, 
the Fe-peak elements extend further out in Lagrangian mass coordinate (see Figure \ref{fig:Chemprof}).
This translates into very different shocked masses of each element at a given age and 
ambient medium density for different explosion models, and therefore into large differences 
in the X-ray spectra. 
For Si and S, on the other hand, most of the ejected mass is already shocked at 538 years 
in all models ($M_{\rm{shocked}} = 0.81, 0.90, 0.98, 1.06 \, M_{\odot}$ for models 
SCH088\_2p0, SCH097\_2p0, SCH106\_2p0, SCH115\_2p0, and 
$M_{\rm{shocked}} = 1.16, 1.18, 1.20, 1.21 \, M_{\odot}$ for models DDT12\_2p0, 
DDT16\_2p0, DDT24\_2p0, DDT40\_2p0, shown in Figure \ref{fig:Chemprof}), which translates into 
a smaller dynamic range of X-ray emitting masses and therefore smaller differences for the 
corresponding lines in the spectra. Elements like Mg and O are also fully shocked at this age, 
but their spectral blends show larger variations than those of Si and S because the dynamic 
range in ejected masses is much larger (see Table \ref{table:progenitor_models}).

Our spectral models can also be convolved with the response matrices for future facilities, 
like the X-Ray Imaging and Spectroscopy Mission 
\citep[\textit{XRISM}, a.k.a. X-Ray Astronomy Recovery Mission, \textit{XARM},][]{Tas18} or 
\textit{Athena} \citep{Na13}. The left panel of Figure \ref{fig:Sp_phSuXA} shows the RS and FS spectra for 
model SCH115\_2p0 at 538 years, unconvolved (photon flux) and after convolution with both 
\textit{Suzaku} and \textit{XRISM} responses. It is worth noting that \textit{XRISM} will not be 
able to separate the FS and RS for the remnants in our sample. 
The improved energy resolution of \textit{XRISM} reveals a wealth of transitions that 
cannot be seen with \textit{Suzaku}, as shown in the 
right panel of Figure \ref{fig:Sp_phSuXA}. There are two transitions at
$\approx$ 5.4 and $\approx$ 5.65 keV in both the \textit{Suzaku} and the \textit{XRISM}
synthetic spectrum that do not appear in real \textit{Suzaku} observations. We defer this
to a future study.

The one-dimensional nature of our models deserves some comments. Multidimensional hydrodynamics 
coupled with NEI calculations \citep{War13,Or16} are computationally expensive, and do not allow to produce 
extensive model grids for an exhaustive exploration of parameter space like the one we present here. The 
results from \citet{War13}, who studied the impact of clumping and Rayleigh-Taylor instabilities in the 
morphology and ionization (but not emitted spectra) of Type Ia SNRs in 3D, 
do not show major deviations from one-dimensional calculations.

\section{Discussion}\label{sec:discussion}

\subsection{Type Ia SNRs: Bulk properties}\label{subsec:bulk}

Here we describe the bulk properties (expansion age, radius, Fe K$\alpha$ centroid, and 
Fe K$\alpha$ luminosity) of our \mch\ and sub-\mch\ models and compare them with the available 
observational data for Ia SNRs. We use the Fe K$\alpha$ blend because it is sensitive to the 
electron temperature and ionization timescale in SNRs, with the centroid energy being a strong 
function of mean charge state \citep{Vi12,Ya14a,Ya14b}. This results in a clear division between 
Ia SNRs, which tend to interact with a low-density ambient medium, and core collapse (CC) SNRs, 
which often evolve in the high density CSM left behind by their massive and short-lived 
progenitors (first noted by \citealt{Ya14b}, see also \citealt{Pat15,PatB17}). In their analysis, 
\citet{Ya14a} already found that the bulk properties of the SNRs identified as Ia in their sample 
(those with Fe K$\alpha$ centroid energies below 6.55 keV) were well reproduced by the \mch\, 
uniform ambient medium models of \citet{Ba03,Ba05}. Here, we perform a more detailed comparison 
to our models, which also assume a uniform ambient medium, but are based on an updated code and 
atomic data, and include both \mch\ and sub-\mch\ progenitors. We also comment briefly on some 
individual objects of interest.

We calculate the Fe K$\alpha$ centroid energy $E_{\, \rm{Fe}_{K\alpha}}$ and luminosity 
$L_{\, \rm{Fe}_{K\alpha}}$ for each model as

\begin{equation}
E_{\, \rm{Fe}_{K\alpha}} =  \dfrac{ \mathlarger{\int_{E_{\rm{min}}}^{E_{\rm{max}}} \left( F \times \, E \right) \, dE} } { \mathlarger{\int_{E_{\rm{min}}}^{E_{\rm{max}}} F  \, dE} }
= \dfrac{ \mathlarger{\sum_{i \subseteq}^{}} \, F_{i} \times E_{i} \times dE_{i} } {  \mathlarger{\sum_{i \subseteq}^{}} \, F_{i} \times dE_{i} }
\end{equation}
\vspace{-0.3cm}
\begin{equation}
F_{\, \rm{Fe}_{K\alpha}}  = \mathlarger{\int_{E_{\rm{min}}}^{E_{\rm{max}}}} F  \, dE = \mathlarger{\sum_{i \subseteq}^{}} \, F_{i} \times dE_{i}
\end{equation}
\vspace{-0.3cm}
\begin{equation}
L_{\, \rm{Fe}_{K\alpha}} = 4 \pi D[\textrm{cm}]^{2} \times \, F_{\, \rm{Fe}_{K\alpha}}
\end{equation}

where $F$ is the differential flux from the nonconvolved spectrum after continuum subtraction, 
$dE$ is the constant (1.2 eV) energy step, and $E_{\rm{min}} - E_{\rm{max}}$ is an energy interval 
that covers the entire Fe K$\alpha$ complex (6.3 $-$ 6.9 keV). We only compute these numbers 
when the Fe K$\alpha$ emission is clearly above the continuum.

\begin{figure*}
\centering
\hspace{-1.1 cm}
\includegraphics[scale=0.31]{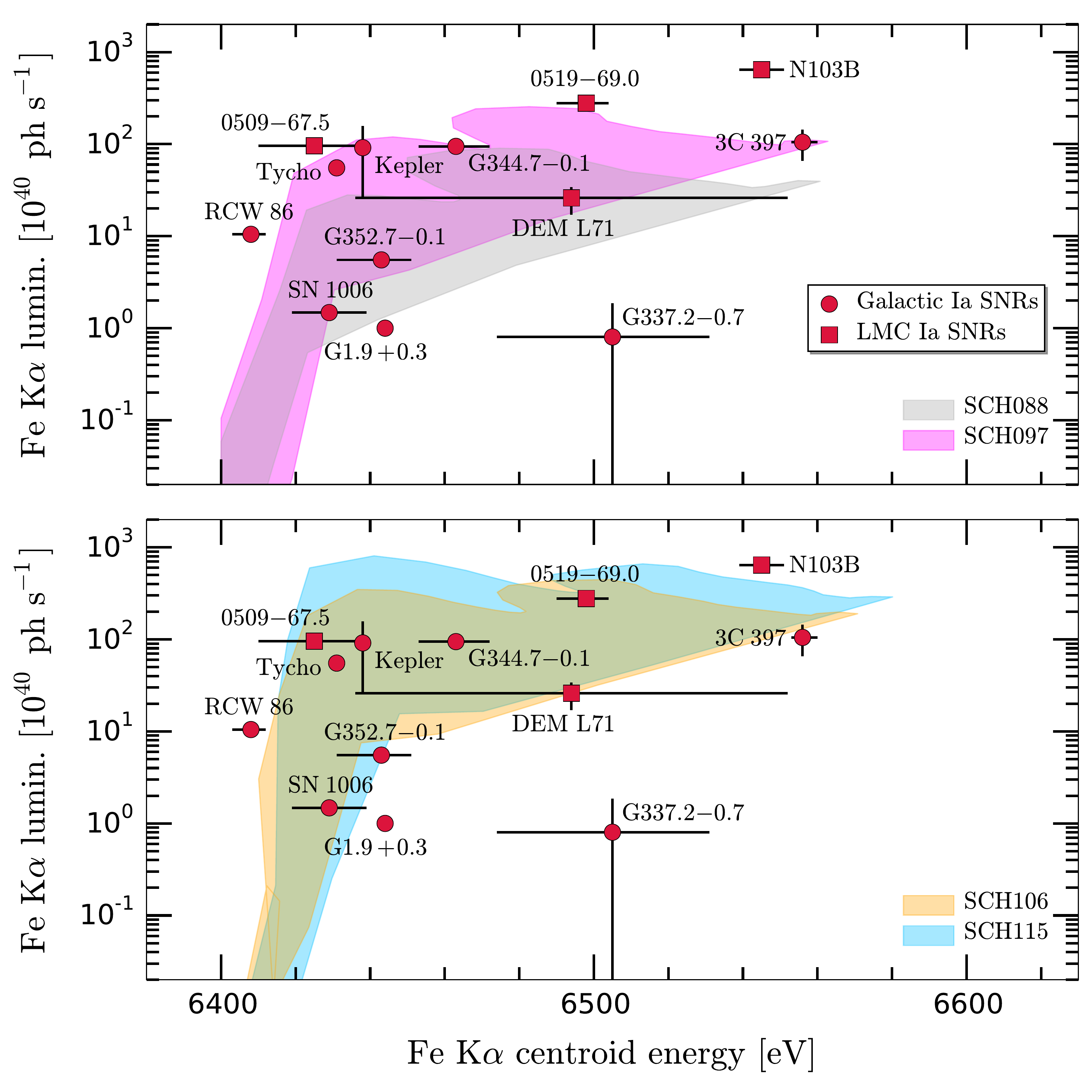}
\includegraphics[scale=0.31]{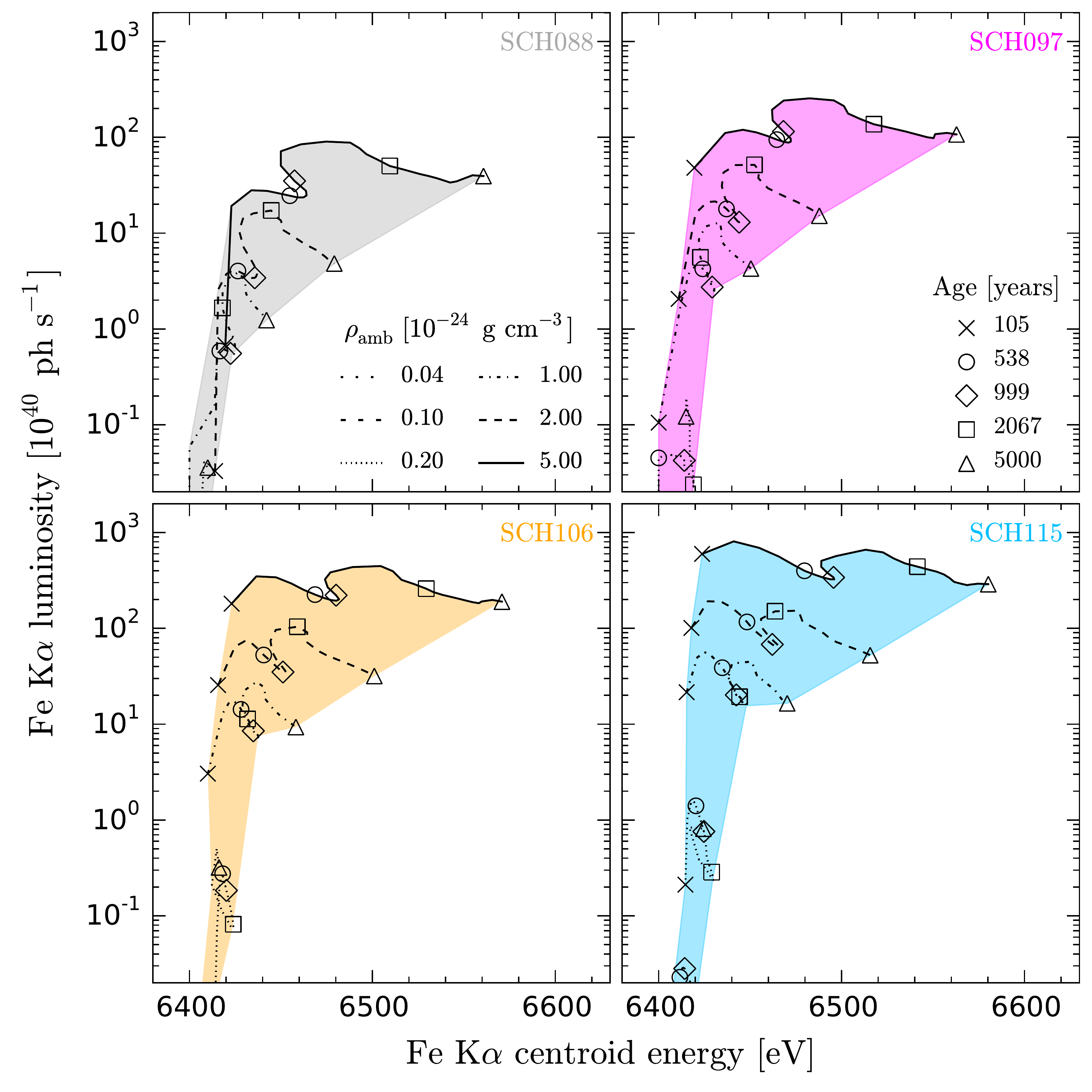}
\vspace{1 cm}
\hspace{-1.1 cm}
\includegraphics[scale=0.31]{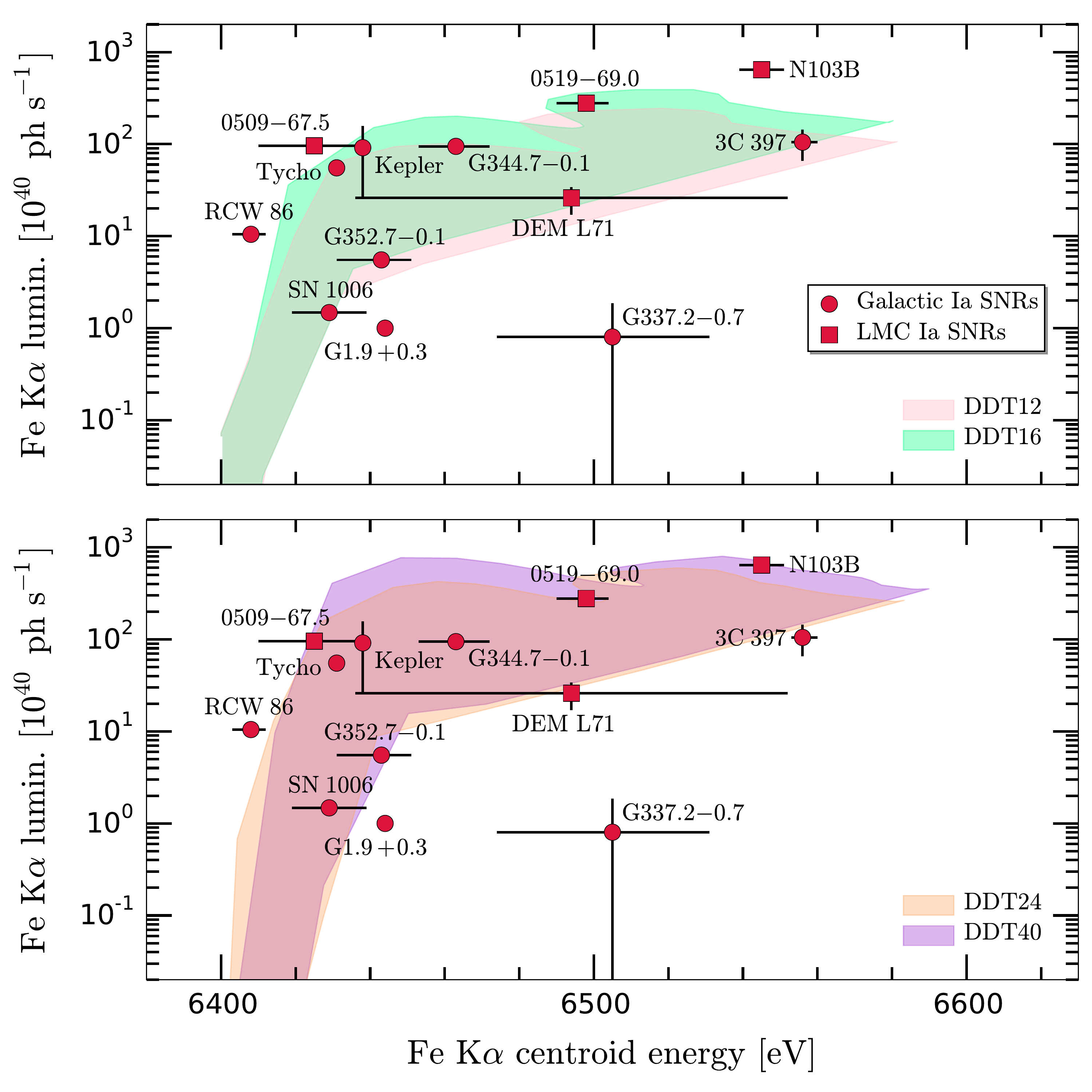}
\includegraphics[scale=0.31]{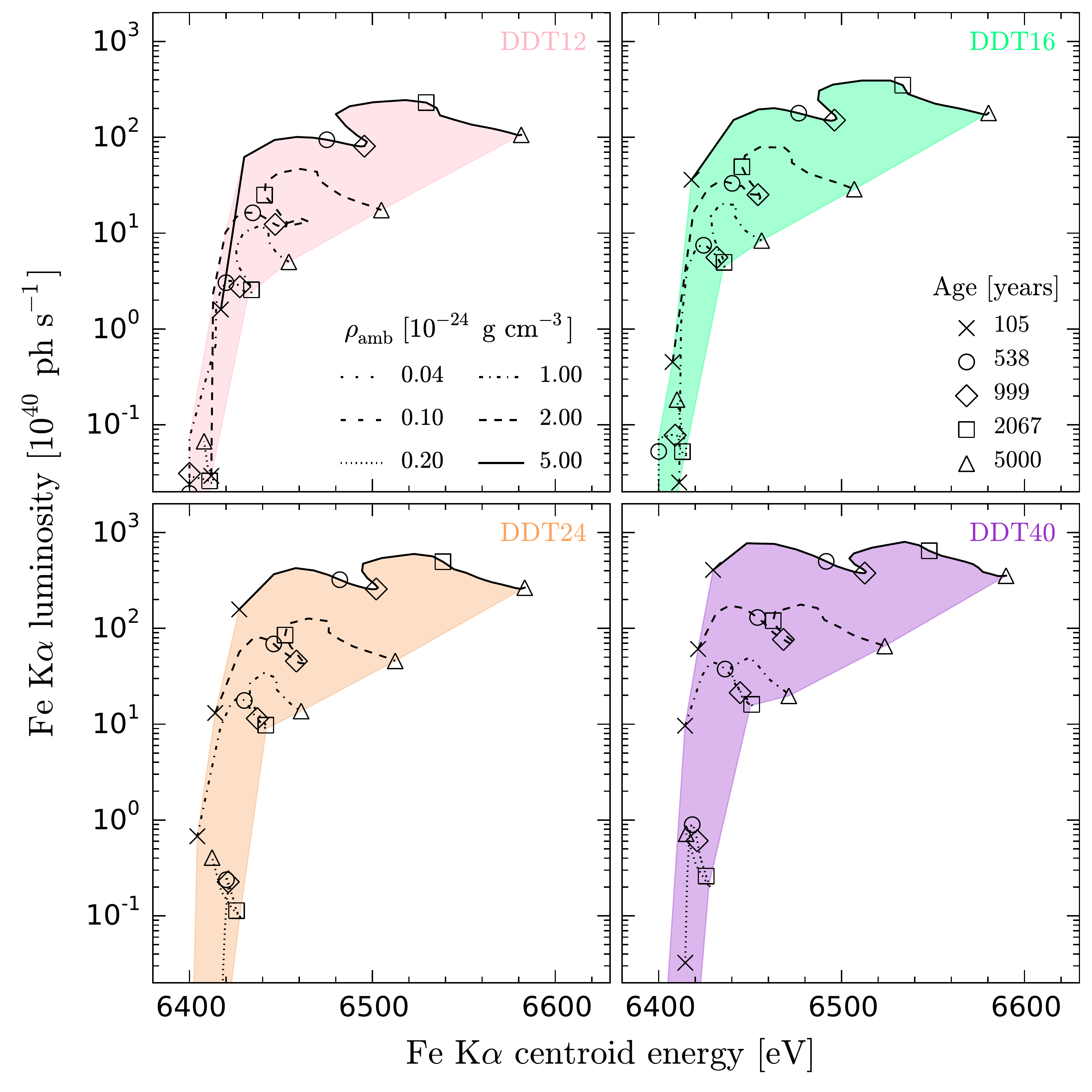}
\caption{Left: Centroid energies and line luminosities of Fe K$\alpha$ emission from various Type Ia
SNRs in our Galaxy (circles) and the LMC (squares). The shaded regions depict the Fe 
K$\alpha$ centroids and luminosities predicted by our theoretical sub-\mch\ and \mch\ models
with various uniform ISM densities (SCH088: gray; SCH097: magenta; SCH106: orange;
SCH115: blue; DDT12: pink; DDT16: green; DDT24: light brown; DDT40: purple). Right:
Individual tracks for each model. The
$L_{\, \rm{Fe}_{K\alpha}}$ $-$ $E_{\, \rm{Fe}_{K\alpha}}$ tracks corresponding to the 
two lowest ambient densities ($\rho_{0p04}$, $\rho_{0p1}$) do not appear in the plots
because their $L_{\, \rm{Fe}_{K\alpha}}$ values are considerably small.}
\label{fig:LvsE}
\end{figure*}

\begin{figure*}
\centering
\hspace{-1.1 cm}
\includegraphics[scale=0.31]{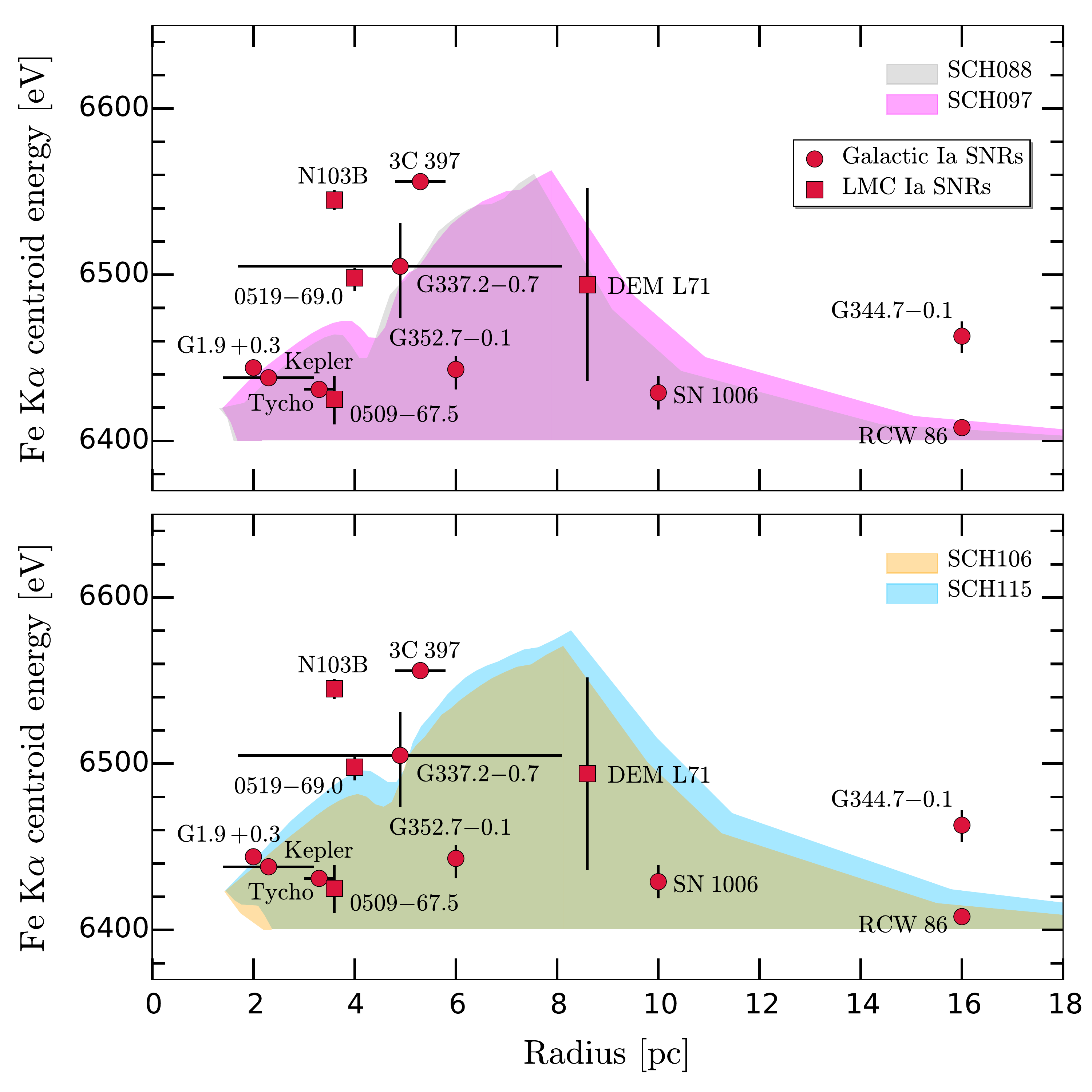}
\includegraphics[scale=0.31]{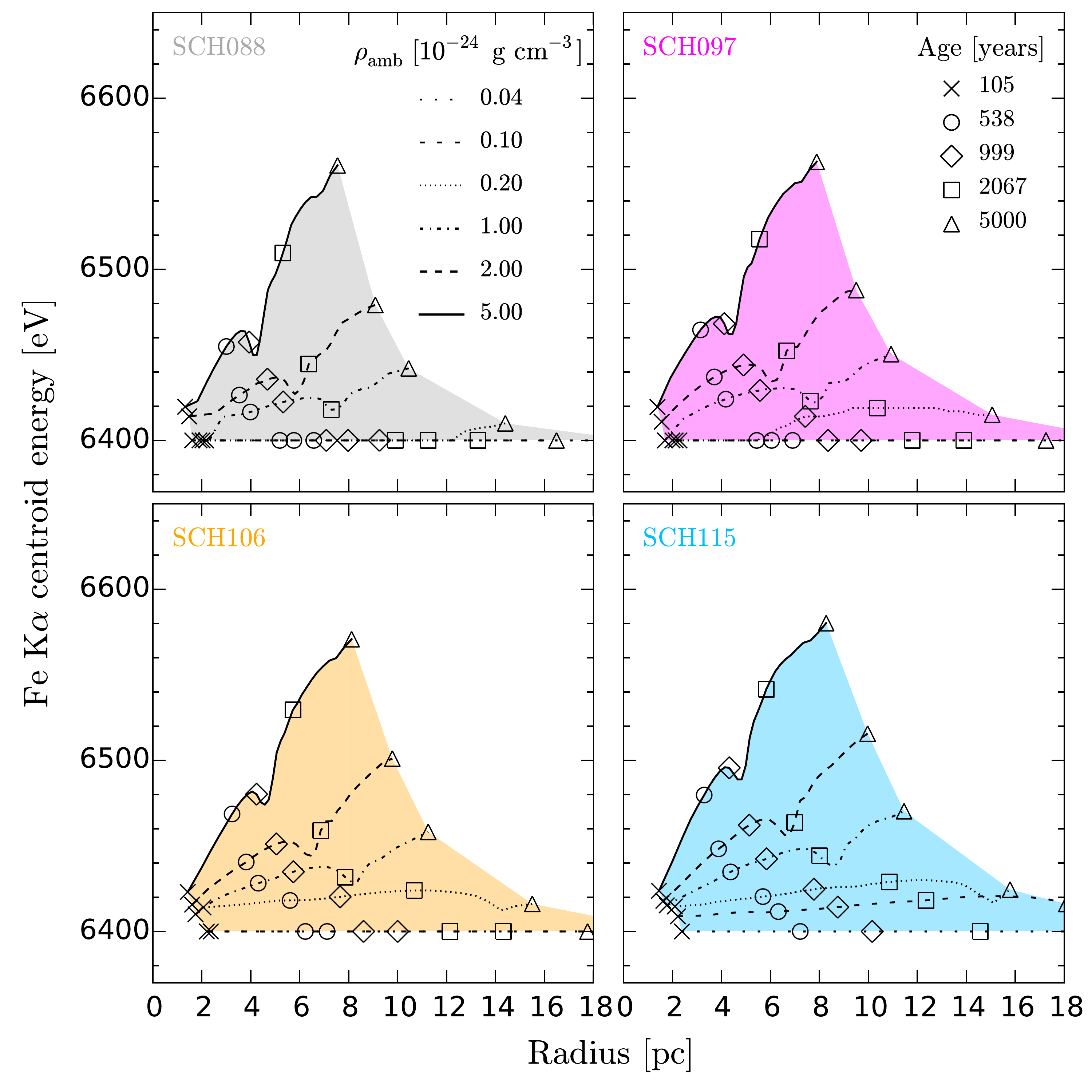}
\vspace{1 cm}
\hspace{-1.1 cm}
\includegraphics[scale=0.31]{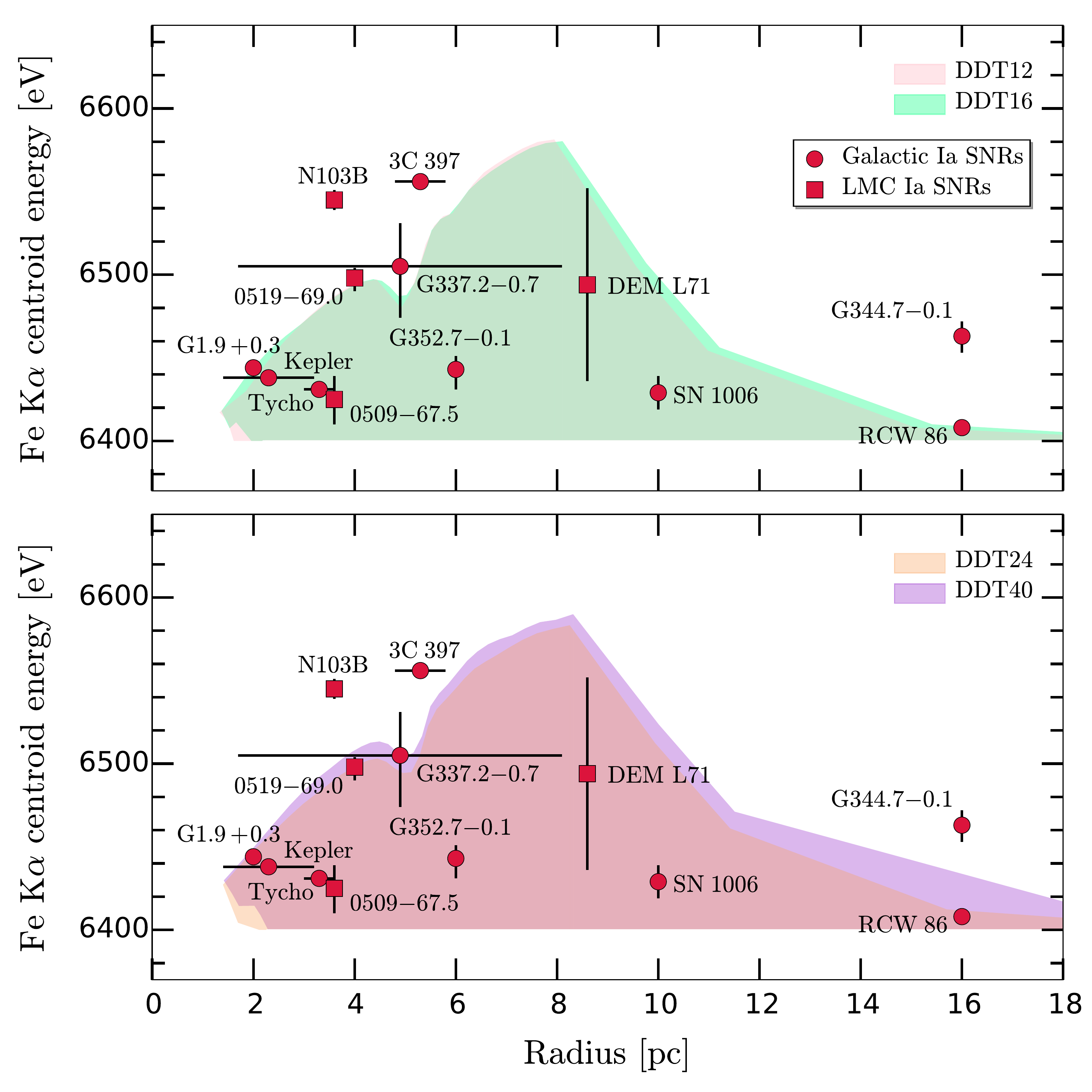}
\includegraphics[scale=0.31]{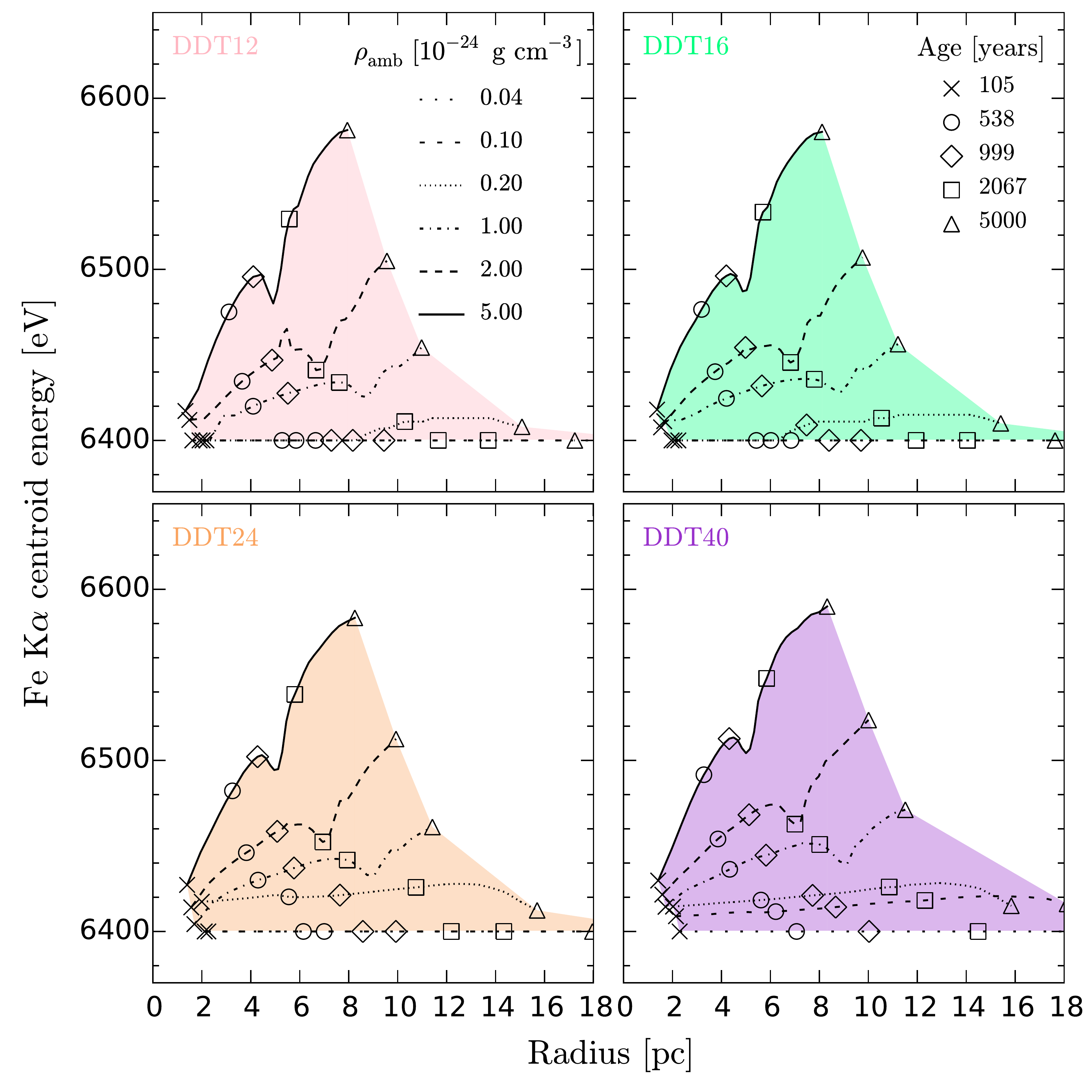}
\caption{Fe K$\alpha$ centroid energy versus forward shock radius for the Type Ia SNRs 
in our sample. The shaded regions correspond to the models shown in Figure 
\ref{fig:LvsE}.}
\label{fig:EvsR}
\end{figure*}

\begin{figure*}
\centering
\hspace{-1.1 cm}
\includegraphics[scale=0.31]{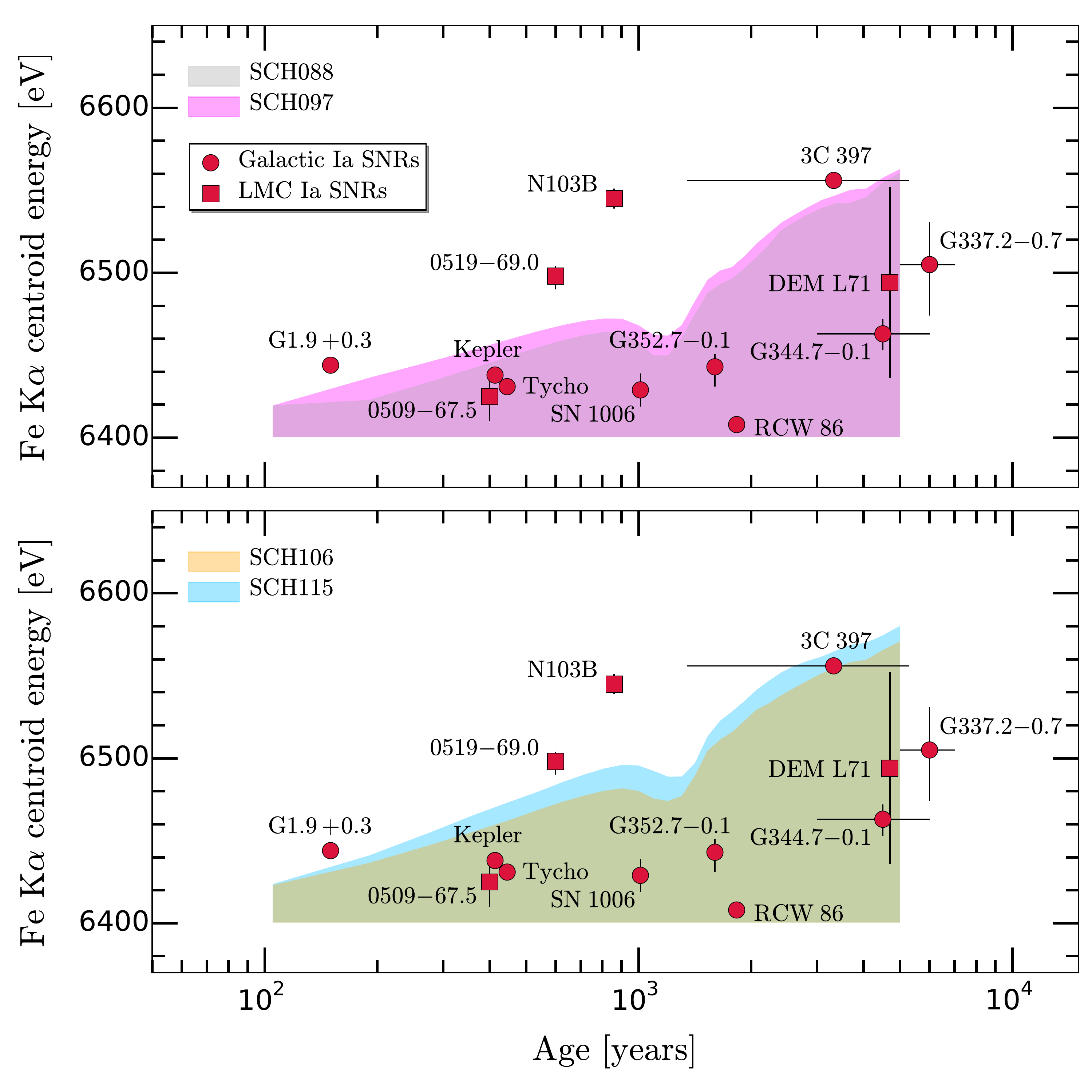}
\includegraphics[scale=0.31]{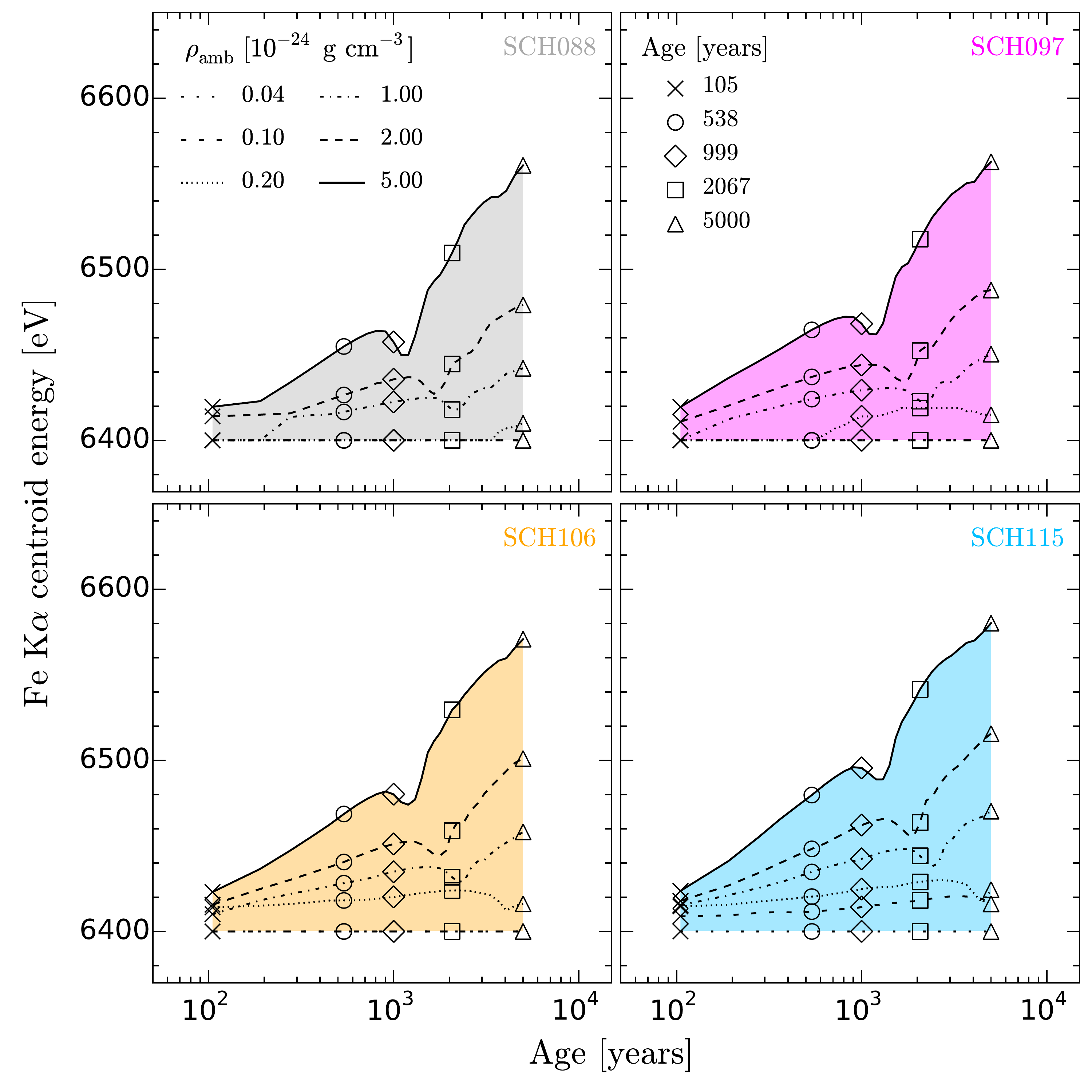}
\vspace{1 cm}
\hspace{-1.1 cm}
\includegraphics[scale=0.31]{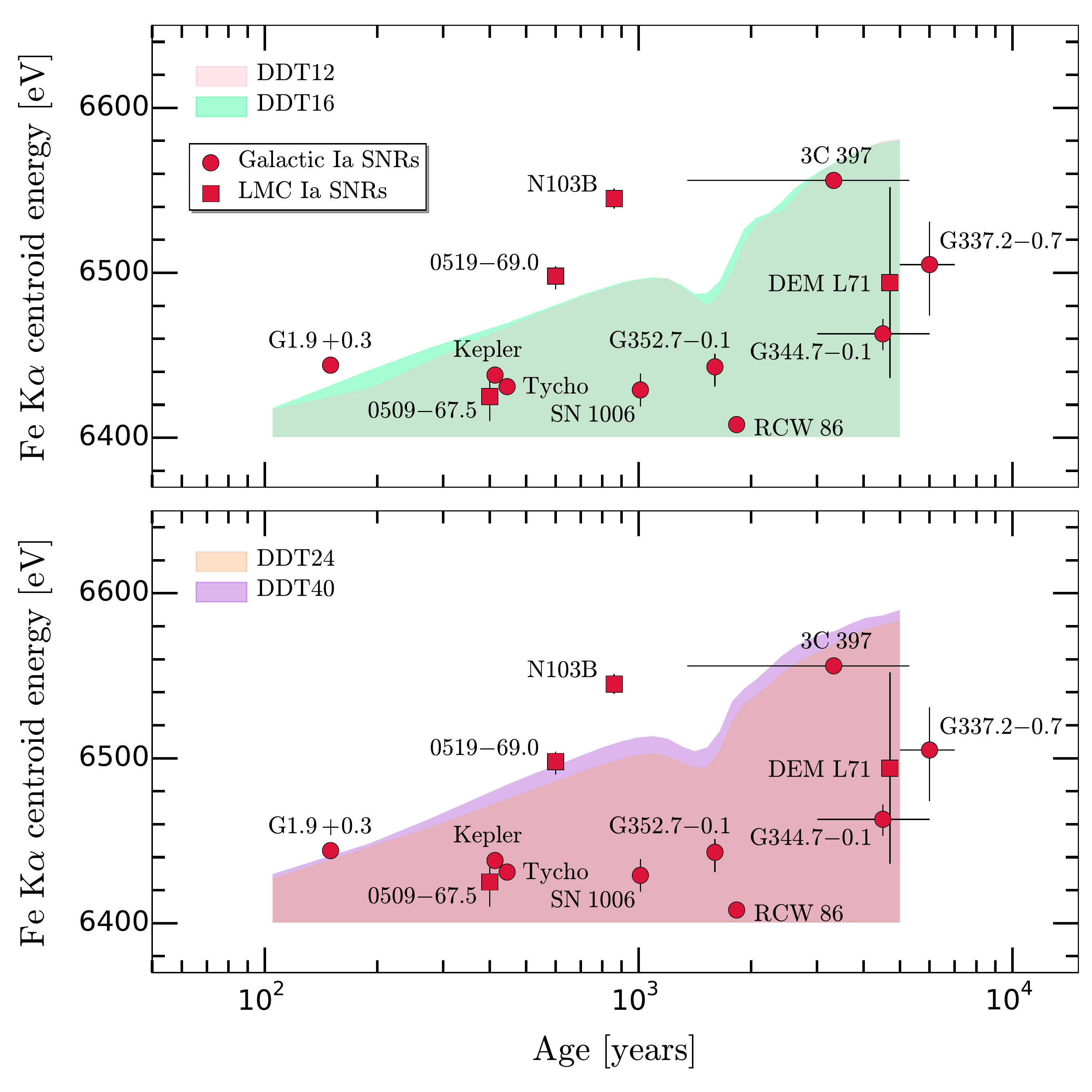}
\includegraphics[scale=0.31]{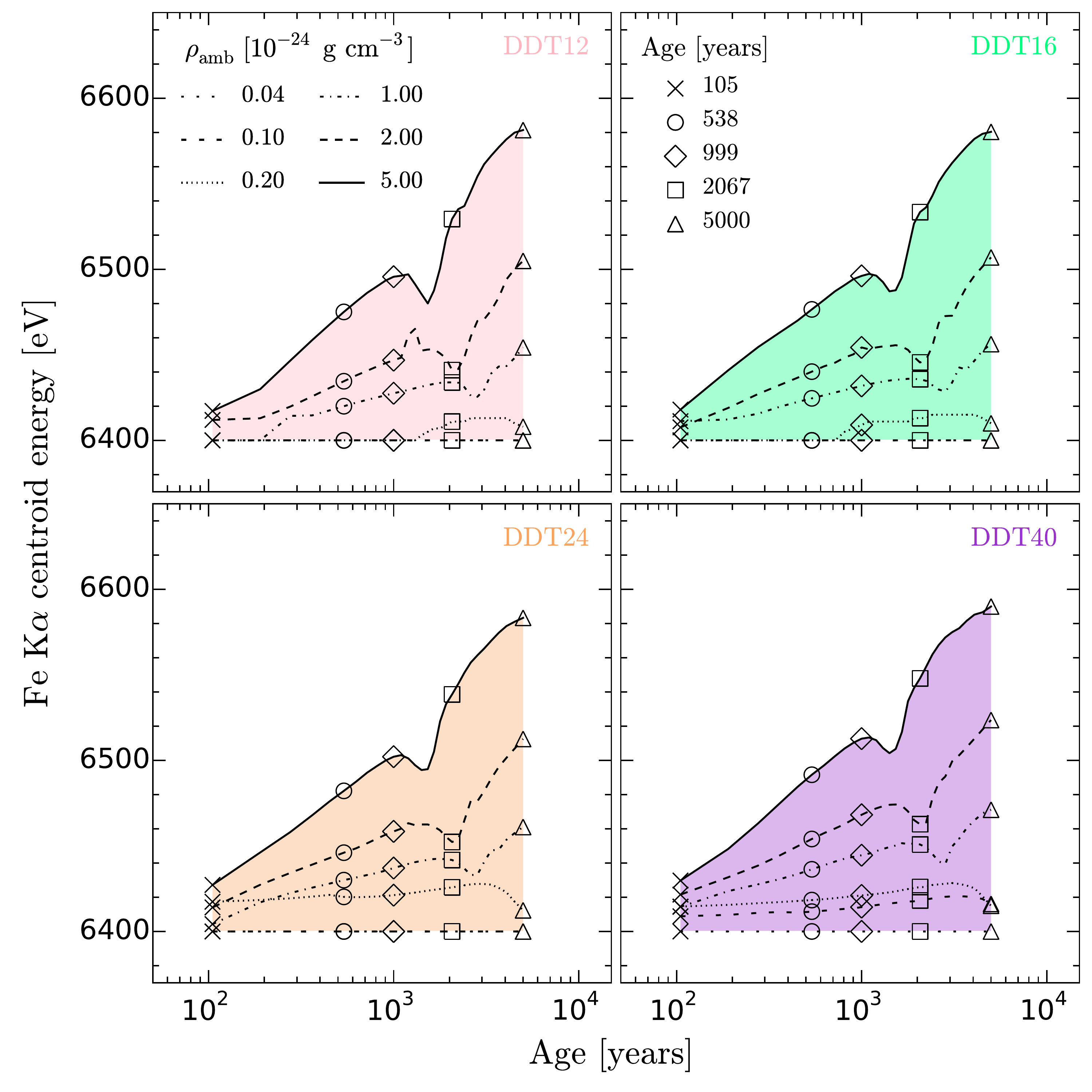}
\caption{Fe K$\alpha$ centroid energy versus expansion age for the Type Ia SNRs in 
our sample. The shaded regions correspond to the models shown in Figures \ref{fig:LvsE}
and \ref{fig:EvsR}.}
\label{fig:Evst}
\end{figure*}

\begin{figure*}
\centering
\hspace{-1.1 cm}
\includegraphics[scale=0.31]{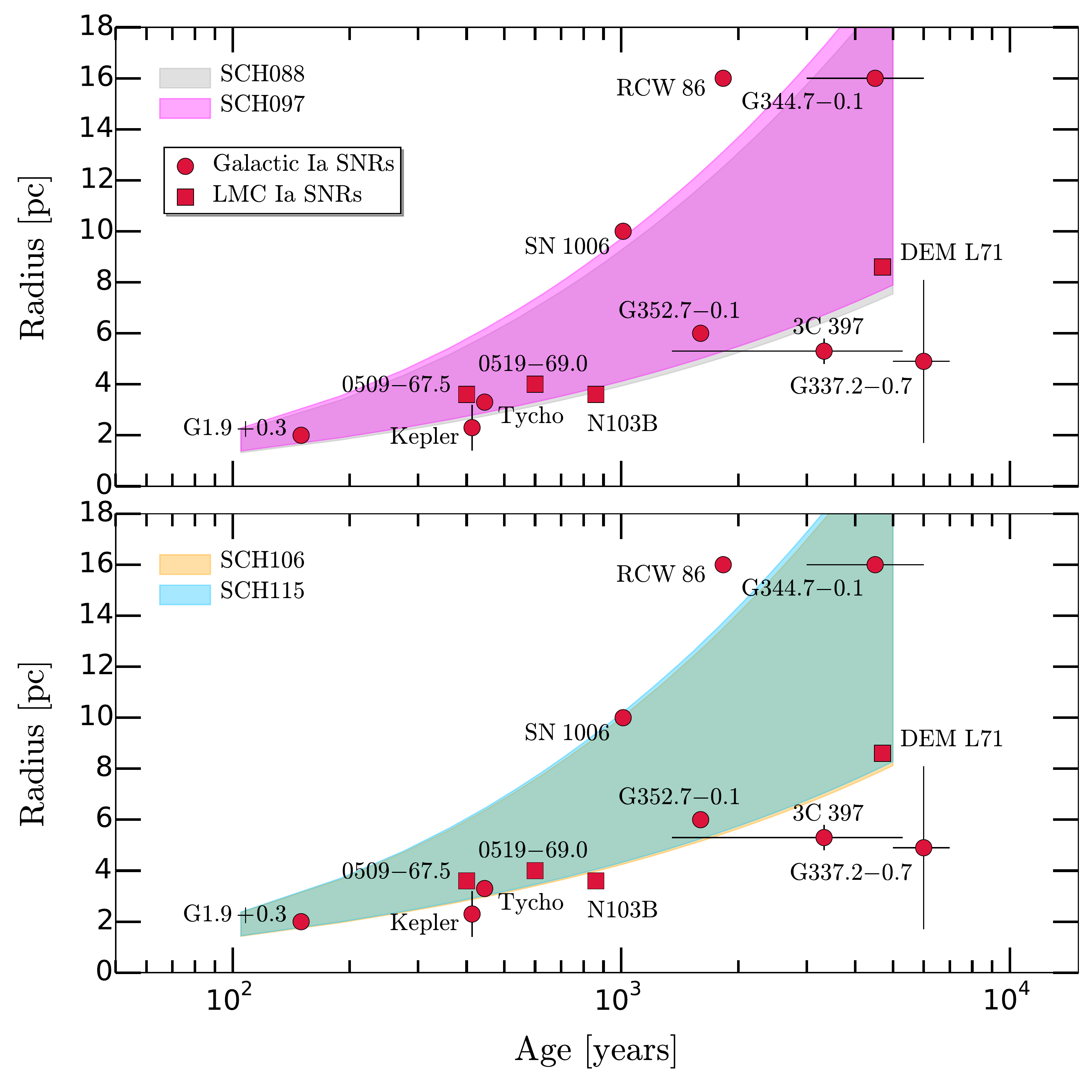}
\includegraphics[scale=0.31]{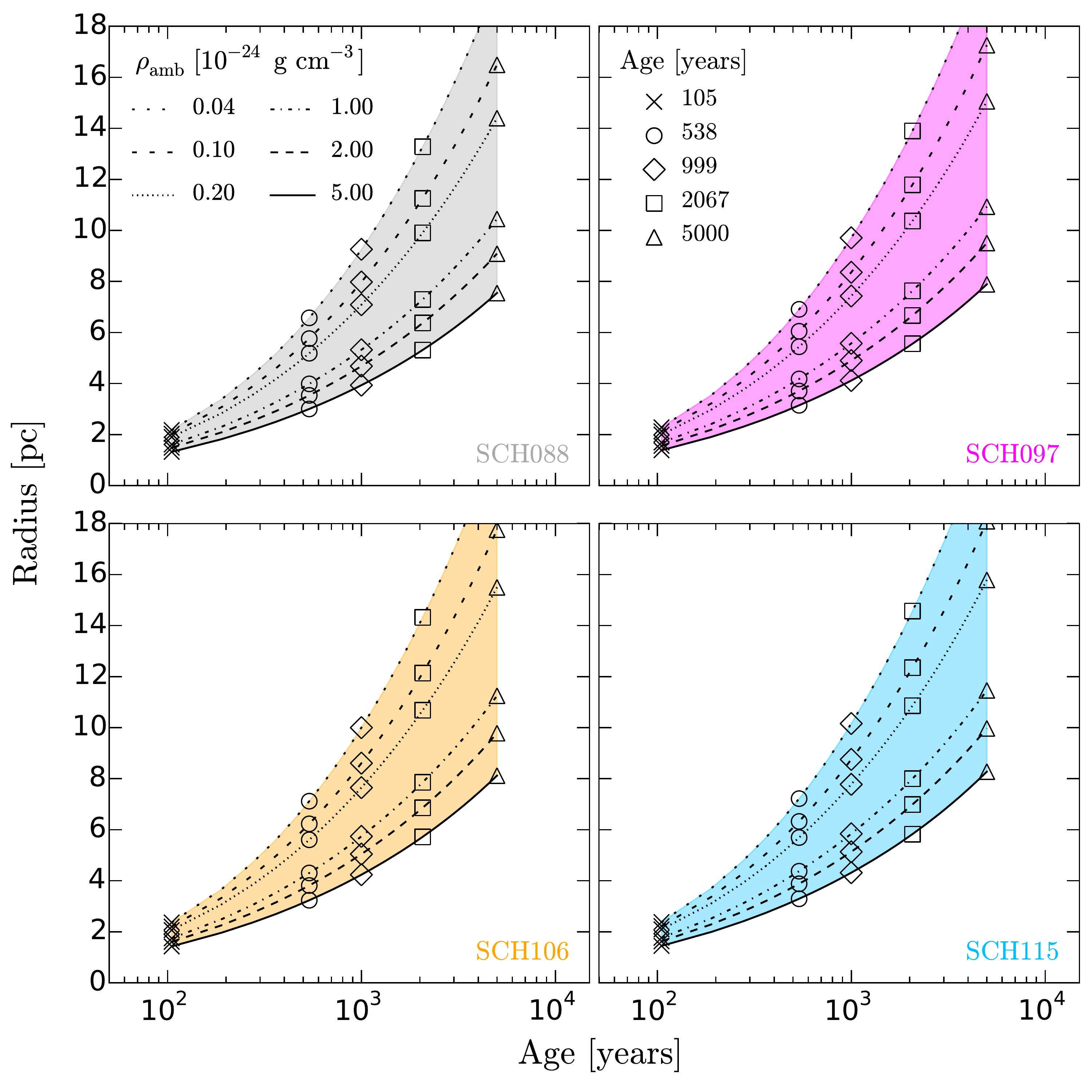}
\vspace{1 cm}
\hspace{-1.1 cm}
\includegraphics[scale=0.31]{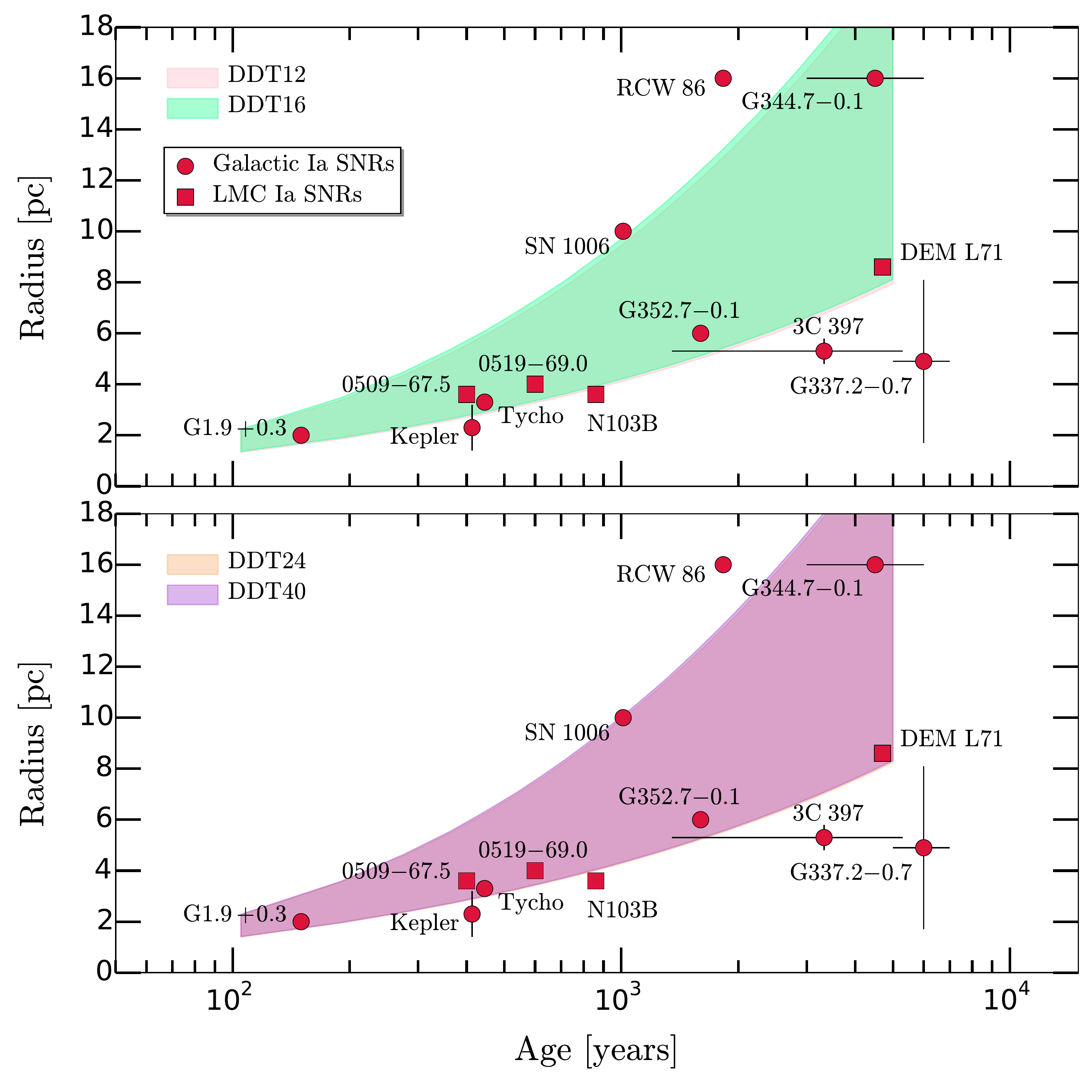}
\includegraphics[scale=0.31]{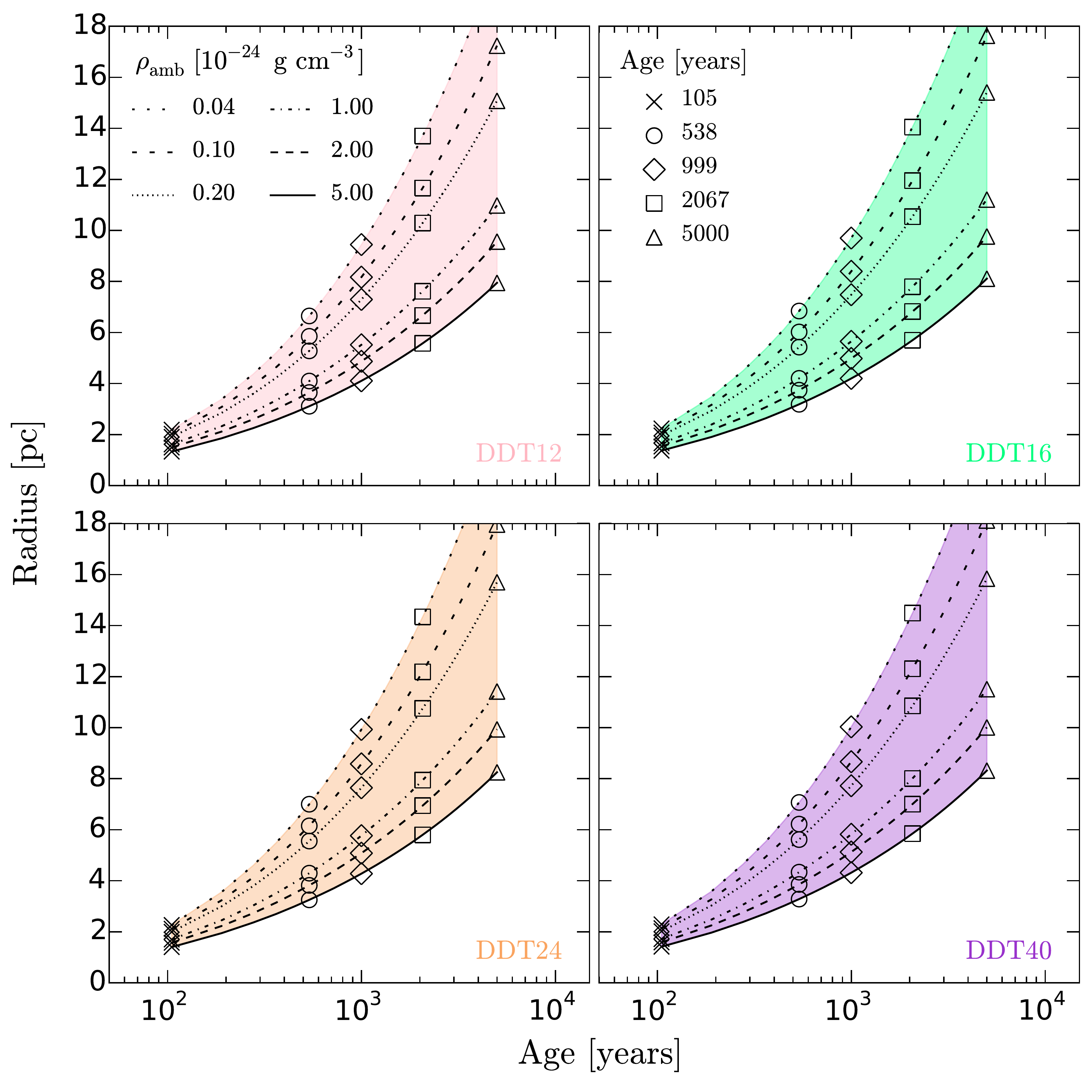}
\caption{Forward shock radius versus expansion age for the Type Ia SNRs in our sample. The shaded 
regions correspond to the models shown in Figures \ref{fig:LvsE}, \ref{fig:EvsR}
and \ref{fig:Evst}.}
\label{fig:Rvst}
\end{figure*}

Table \ref{table:remmants} summarizes the relevant observational properties of the 13 Type Ia SNRs in our 
sample. The data are taken from \citet{Ya14a} (\textit{Suzaku} observations). We also include the 
\textit{Chandra} measurements for G1.9+0.3 \citep{Bo13} and the \textit{XMM--Newton} results for
DEM L71 \citep{Mag16}. The contours in Figures \ref{fig:LvsE}$-$\ref{fig:Rvst} show the parameter 
space spanned by our models, with symbols indicating the observed properties of individual SNRs. 
We display $L_{\, \rm{Fe}_{K\alpha}}$ versus $E_{\, \rm{Fe}_{K\alpha}}$ (Figure \ref{fig:LvsE}), 
$E_{\, \rm{Fe}_{K\alpha}}$ versus FS radius ($R_{\rm{FS}}$, Figure \ref{fig:EvsR}), 
$E_{\, \rm{Fe}_{K\alpha}}$ versus expansion age (Figure \ref{fig:Evst}), 
and $R_{\rm{FS}}$ versus expansion age (Figure \ref{fig:Rvst}).

The main features of the models shown in these plots merit some comments. In Figures 
\ref{fig:LvsE}$-$\ref{fig:Evst}, for the models with $\rho_{1p0}$, $\rho_{2p0}$ and 
$\rho_{5p0}$, $E_{\, \rm{Fe}_{K\alpha}}$ decreases for a short time
$\approx 1000-2000$ years after the explosion instead of increasing monotonically with time. 
This is due to the reheating of the shocked ejecta after the RS bounces at the SNR center. 
The reshocked material becomes denser and hotter, and therefore more luminous. This results in 
a lower luminosity-weighted ionization state for the shocked ejecta, which prior to RS bounce 
was dominated by the dense, highly ionized material close to the CD. As time goes on and the 
entire ejecta is reshocked, the material close to the CD dominates the spectrum again, 
and the ionization state continues to increase monotonically. 
The strength of this feature is due to the spherical symmetry of our models, at least to some 
extent, but we expect a qualitatively similar (if weaker) effect in reality. We note that, 
although our model predictions are qualitatively similar to those from \citet{Ba03,Ba05,Ba06}, 
\citet{Ya14a} and \citet{Pat15}, there are small deviations $-$ for instance, we predict a 
slightly higher $E_{\, \rm{Fe}_{K\alpha}}$ for the same ambient medium density and age 
(${\sim}$ 6.6 keV versus ${\sim}$ 6.5 keV). This is likely due to differences in the hydrodynamic code, 
atomic data, and explosion models. In addition, \citet{Pat15} stopped their calculations when
the RS first reached the center of the SNR, while we continue ours until the models reach an age 
of 5000 years.

\placefigure{f9}

\placefigure{f10}

\placefigure{f11}

\placefigure{f12}

Figures \ref{fig:LvsE}$-$\ref{fig:Rvst} show that the parameter space covered by 
our spherically symmetric, uniform ambient medium models is in good agreement with 
the observed data. While there are exceptions, which we discuss in detail below, 
it is clear that our models are a good first approximation to interpret the bulk 
dynamics of real Type Ia SNRs, and can be used to infer their fundamental physical 
properties. For example, denser ambient media and more energetic progenitor models 
predict higher $E_{\, \rm{Fe}_{K\alpha}}$ and $L_{\, \rm{Fe}_{K\alpha}}$ at a given 
expansion age, as seen in Figure \ref{fig:LvsE}. Thus, the SNRs with the highest 
$L_{\, \rm{Fe}_{K\alpha}}$, like 0519$-$69.0 and 0509$-$67.5, are only compatible 
with the brightest, most Fe-rich progenitor models (SCH106, SCH115, DDT16, and DDT24). 
The Fe K$\alpha$ emission from SNR N103B, in particular, can only be reproduced by 
model DDT40 at the highest ambient medium density. As shown in Figures \ref{fig:EvsR} 
and \ref{fig:Rvst}, $R_{\rm{FS}}$ has a weak dependence on the ejecta mass, 
but it is quite sensitive to the ambient density because 
$R_{\rm{FS}} \propto M^{1/3} \rho^{-1/3}$ \citep{McK95}. Therefore, objects 
surrounded by low-density media (e.g. RCW 86, SN 1006, and G344.7$-$0.1) clearly 
stand apart from those evolving in high density media (e.g. 3C 397, N103B, and 
Kepler): the former have large $R_{\rm{FS}}$ and low $E_{\, \rm{Fe}_{K\alpha}}$ 
centroids, while the latter have small $R_{\rm{FS}}$ and high $E_{\, \rm{Fe}_{K\alpha}}$.
We note that the ages of these remnants differ from one another.
In general, the densities we infer from simple comparisons to our models are in good 
agreement with detailed studies of individual objects. For instance, \citet{So14} 
and \cite{Wi14} determined $n_{\rm{amb}} \gtrsim 2.0 \, \rm{cm^{-3}}$ for N103B, and 
\citet{Lea16} found $n_{\rm{amb}} \ {\sim} \ 2-5 \, \rm{cm}^{-3}$ for 3C 397, which 
are close to the highest value of $\rho_{\rm{amb}}$ in our grid 
($n_{\rm{amb}} = 3.01 \, \rm{cm^{-3}}$).

For all the observables shown in Figures \ref{fig:LvsE}$-$\ref{fig:Rvst}, the main 
sources of variation in the models are the ambient density and the expansion age. 
This implies that the details of the energetics and chemical composition in the 
supernova model, and in particular whether the progenitor was \mch\ or sub-\mch, 
are not the main drivers for the bulk dynamics of Type Ia SNRs. This does not imply 
that our SNR models do not have the power to discriminate Type Ia SN explosion 
properties - detailed fits to the X-ray spectra of individual objects have shown 
that they can do this very well \citep[e.g.,][]{Ba06,Ba08a,Pat12}. However, the bulk 
SNR properties \textit{on their own} are not very sensitive to the explosion 
properties, especially for objects whose expansion ages or distances are not well 
determined. To discriminate explosion properties, additional information needs to 
be taken into account, like specific line flux ratios 
(e.g. $\rm{Si \,\, K\alpha \, / \, Fe \,\, K\alpha}$, 
$\rm{S \,\, K\alpha \, / \, Fe \,\, K\alpha}$, and 
$\rm{Ar \,\, K\alpha \, / \, Fe \,\, K\alpha}$), which can distinguish \mch\ 
from sub-\mch\ progenitors, or even better, detailed fits 
to the entire X-ray spectrum, which can reveal a wealth of information about the 
explosion \citep[e.g.,][]{Ba06,Ba08a,Pat12}. We defer these applications of our models 
to future work.

To evaluate the degree to which a particular model works well for a given SNR, it is 
important to examine \textit{all} its bulk properties at the same time. By doing this, 
we can single out individual objects whose bulk dynamics cannot be reproduced by our 
models, modulo any uncertainties in the expansion age and distance. Not surprisingly, 
the SNR that shows the largest deviation from our models is RCW 86. This remnant is 
known to be expanding into a low-density cavity, presumably excavated by a fast, 
sustained outflow from the SN progenitor \citep{Ba07,Wi11a,Bro14}, and therefore its 
$R_{\rm{FS}}$ is too large for its expansion age and $E_{\, \rm{Fe}_{K\alpha}}$. 
In addition, its classification as a Type Ia SNR is still under debate \citep{Gva17}. 
The Galactic SNR G344.7$-$0.1 also shows a similar deviation, 
albeit less strong, but this might be related to an overestimated distance and $R_{\rm{FS}}$ 
\citep[][and references therein]{Ya12b}. 

Among the objects interacting with low-density 
media, the size of SN 1006 is compatible with our lowest-density models, which agrees with
the value  $n_{\rm{amb}} \ {\sim} \ 0.03 \, \rm{cm}^{-3}$ found by \citet{Ya08}, and its
$E_{\, \rm{Fe}_{K\alpha}}$ and $L_{\, \rm{Fe}_{K\alpha}}$ are within the parameter space 
covered by the models. We examine the case of SN 1006 in more detail in Section 
\ref{subsec:historical}. Among the objects interacting with high density media, 3C 397 
and N103B have $E_{\, \rm{Fe}_{K\alpha}}$ values that are too high for their physical 
sizes and expansion ages. This has been pointed out by \citet{PatB17}, and could be due 
to some sort of interaction with dense material, possibly (but not necessarily) a CSM 
modified by the SN progenitor \citep{Sa05,Wi14,Li17}. Remarkably, the bulk dynamics of the 
Kepler SNR, which is often invoked as an example of CSM interaction in Type Ia SNRs 
\citep[e.g.,][]{Rey07,Chi12,Bu13} are compatible with a uniform ambient medium interaction, 
although a detailed spectral analysis suggests the presence of a small cavity
around its progenitor system \citep{Pat12}. Finally, the 
Galactic SNR G337.2$-$0.7 appears to be underluminous for its relatively high 
$E_{\, \rm{Fe}_{K\alpha}}$, but this could be due to the large uncertainty in its distance 
\citep{Ra06}.

\placefigure{f13}
\begin{figure*}
\centering
\includegraphics[scale=0.55]{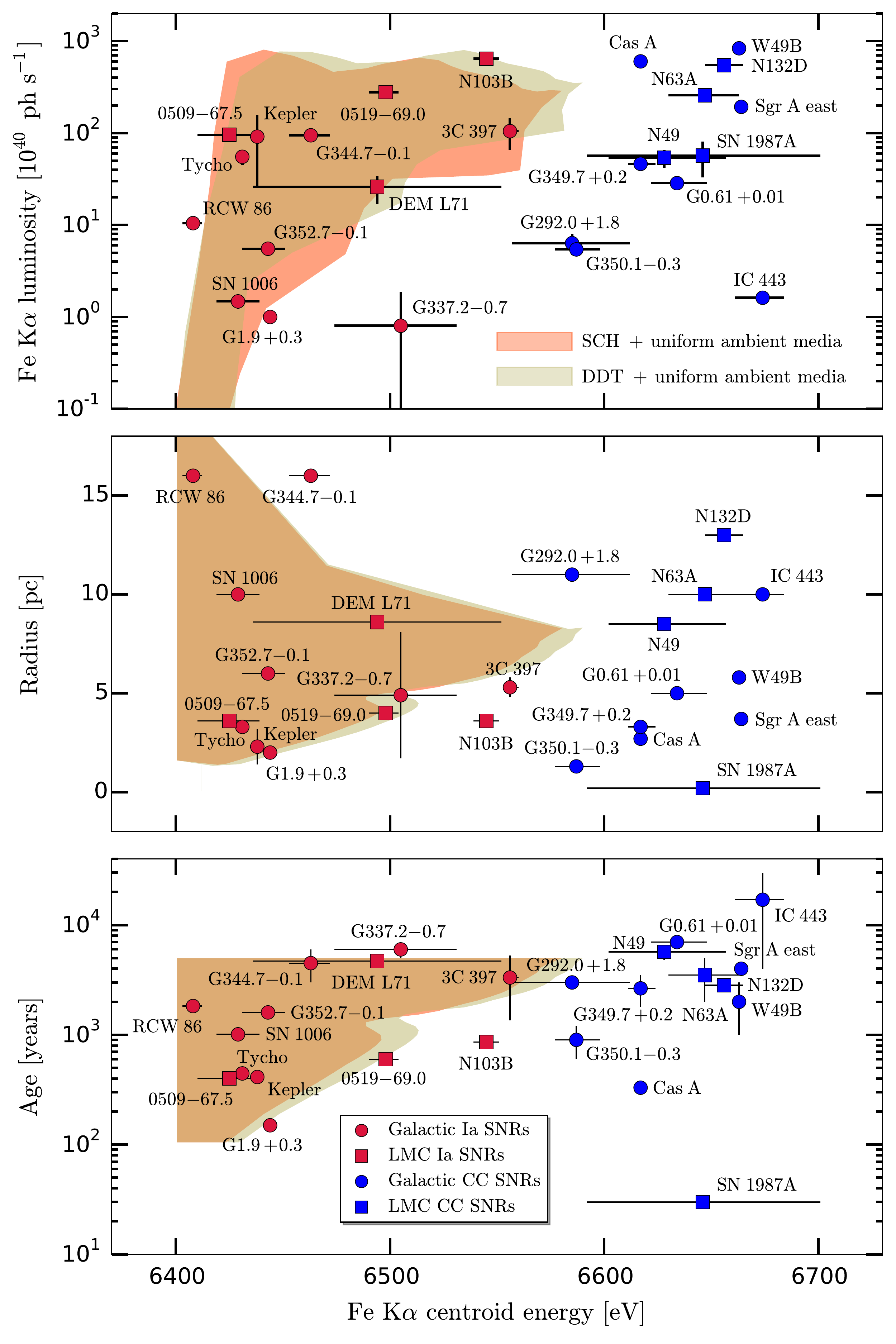}
\caption{Fe K$\alpha$ luminosity, radius and expansion age as a function of the 
Fe K$\alpha$ centroid energy for Ia (red) and CC (blue) SNRs 
\citep[][and references therein]{Lov11,Vo11,Park12,Tia14,Ya14a}.
For a more updated sample and further discussion, see \citet{Mag17}.
The shaded regions depict the predictions from our theoretical 
\mch\ (khaki) and sub-\mch\ (dark orange) models with uniform ISM densities. }
\label{fig:Contours_Ia}
\end{figure*}

\placefigure{f14}
\begin{figure*}
\centering
\includegraphics[scale=0.55]{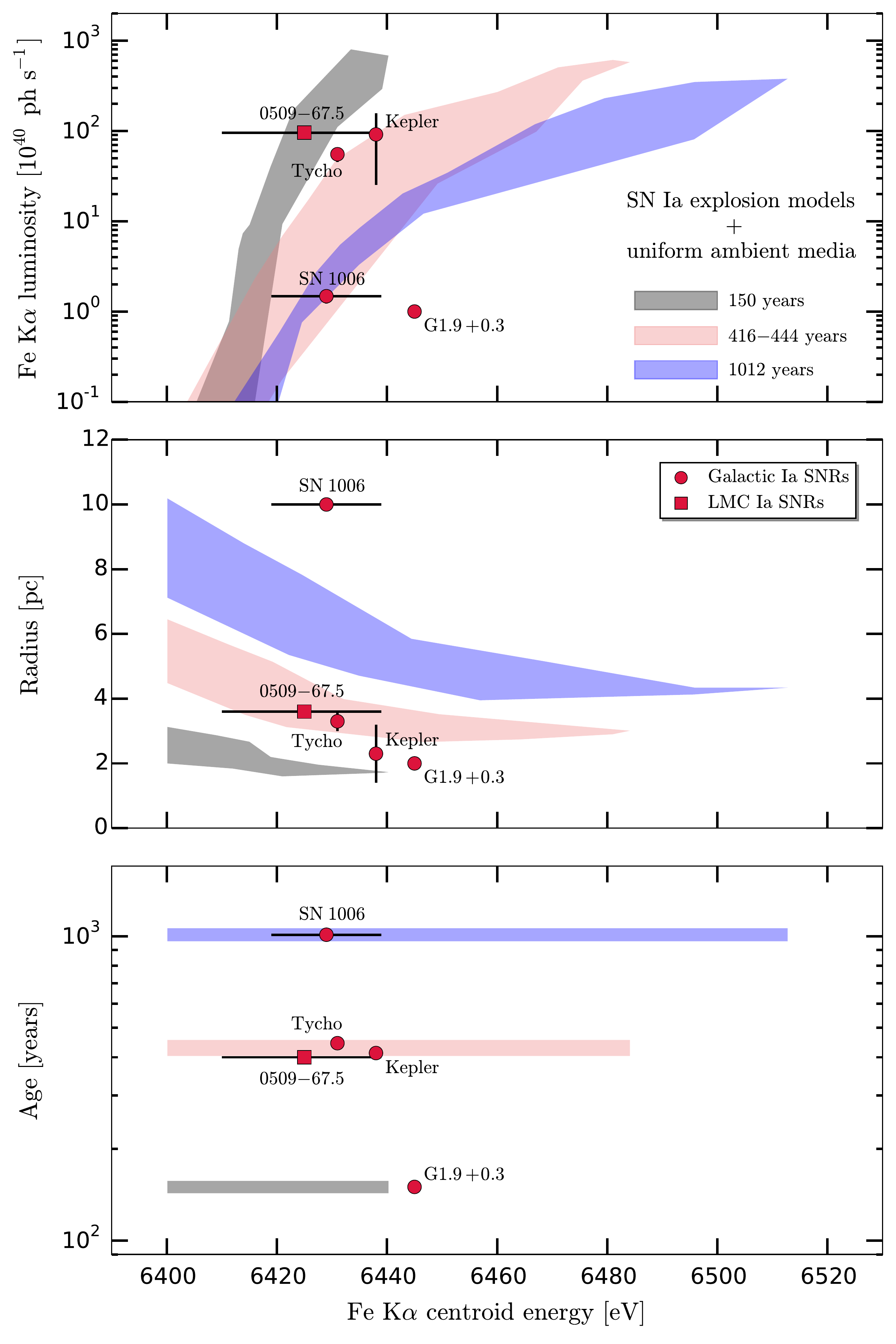}
\caption{Fe K$\alpha$ luminosity, radius and expansion age as a function of the 
Fe K$\alpha$ centroid energy for G1.9+0.3, 0509$-$67.5, Kepler, Tycho, and SN 1006.
The shaded regions depict the predictions from our theoretical 
\mch\ and sub-\mch\ models with uniform ISM densities for different expansion ages:
150 (black), 416$-$444 (light coral), and 1012 (blue) years. }
\label{fig:Historical_contours_Ia}
\end{figure*}

We summarize our comparisons between models and data in Figure \ref{fig:Contours_Ia}, 
which shows $L_{\, \rm{Fe}_{K\alpha}}$, $R_{\rm{FS}}$ and expansion age for our \mch\ 
and sub-\mch\ models and for the SNRs as a function of $E_{\, \rm{Fe}_{K\alpha}}$, the only 
property that can be determined from the observations alone. We re-emphasize that our 
uniform ambient medium, spherically symmetric models, can reproduce the bulk dynamics 
of most Type Ia SNRs quite well. This suggests that, unlike CC SN progenitors, 
most Type Ia SN progenitors do not 
strongly modify their circumstellar environments, as previously noted by 
\citet{Ba07}, \citet{Ya14a}, \citet{PatB17}, and other authors. 
This conclusion is in good agreement with 
the (hitherto unsuccessful) attempts to detect prompt X-ray and radio emission from 
extragalactic Type Ia SNe \citep{Mrg14,Chom16}, but we note that SNR studies probe spatial 
and temporal scales \citep[${\sim}$ pc and ${\sim} \, 10^{5}$ years,][]{PatB17} that are more 
relevant for the pre-SN evolution of Type Ia progenitor models. In this sense, the lack 
of a strongly modified CSM sets Type Ia SNRs clearly apart from CC SNRs \citep{Ya14a}, 
which we also include in Figure \ref{fig:Contours_Ia} for comparison. The only two SNRs
with well-determined properties that are clearly incompatible with our uniform ambient 
medium models are RCW 86 and N103B. These SNRs are probably expanding into some sort 
of modified CSM. In the case of RCW 86, the modification is very strong, and clearly 
due to the formation of a large cavity by the progenitor. In the case of N103B (and 
perhaps also 3C 397), the modification could be due to some dense material left behind 
by the progenitor, but detailed models with nonuniform ambient media are required to 
verify or rule out this claim. In any case, it is clear from Figure \ref{fig:Contours_Ia} 
that the modification of the CSM by the progenitor in N103B must be much weaker than what 
is seen around typical CC SNRs.

\subsection{Type Ia SNRs: Remnants with well-determined expansion ages}\label{subsec:historical}

A reduced subset of Type Ia SNRs have well-determined ages, either because they are 
associated with historical SNe (Kepler, Tycho, and SN 1006 have ages of 414, 446, and 
1012 years, respectively), because they have well-observed light echoes \citep[0509$-$67.5
has an age of ${\sim} \,$ 400 years,][]{Res08}, or because their dynamics put very strong 
constraints on their age 
\citep[G1.9+0.3 has an age of ${\sim} \,$ 150 years,][]{Rey08,Car11,DeH14,Sar17b}. 
These objects are particularly valuable benchmarks for our models, because their known 
ages remove an important source of uncertainty in the interpretation of their bulk 
dynamics.

We perform more detailed comparisons for this set of objects by taking our 
models at 150 years (G1.9+0.3), 416$-$444 years 
(0509$-$67.5, Kepler, and Tycho) and 1012 years (SN 1006). Figure 
\ref{fig:Historical_contours_Ia} shows the same quantities as Figure \ref{fig:Contours_Ia},
but here we display the parameter space covered by our \mch\ and sub-\mch\ models at all 
densities for each of the three age ranges mentioned above. The models at 416$-$444 years 
can reproduce the observed properties of Kepler, Tycho, and 0509$-$67.5 quite well, even 
with the added constraints from the known expansion ages, but we stress that detailed fits 
to the entire X-ray spectra might reveal additional information (see \citealt{Pat12} for 
Kepler, \citealt{Sla14} for Tycho). 
In any case, we can say that the bulk dynamics of these three objects
disfavor variations from a uniform medium interaction as large as those
seen in typical CC SNRs. We note 
that we have made no attempt to quantify the extent of the deviation from a uniform 
ambient medium that could be accommodated while still yielding results that are consistent 
with the observations, as it is beyond the scope of the present work.

For SN 1006, $R_{\rm{FS}}$, $E_{\, \rm{Fe}_{K\alpha}}$, and $L_{\, \rm{Fe}_{K\alpha}}$ 
are well reproduced by our models at 1012 years; though, given its surrounding ambient 
density and physical size, $E_{\, \rm{Fe}_{K\alpha}}$ is larger than can be 
explained by a uniform ambient medium interaction. For G1.9+0.3, $R_{\rm{FS}}$ and 
$L_{\, \rm{Fe}_{K\alpha}}$ are close to the values predicted by our models at 150 years, 
but $E_{\, \rm{Fe}_{K\alpha}}$ is too high to be reconciled with a uniform ambient medium 
interaction. In both cases, the bulk properties of the SNRs might indicate an early interaction 
with some sort of modified CSM. For SN 1006, this might be a low-density cavity, perhaps smaller 
in size than the SNR. For G1.9+0.3, a thin, dense shell that changed the ionization state without 
strongly affecting the dynamics might have been involved, as suggested by \cite{Ch16}. In both 
cases, a detailed exploration of the parameter space for CSM interaction in Type Ia SNRs is 
required to confirm or rule out specific scenarios.’

\section{Conclusions}\label{sec:conclusions}

We have presented a new grid of one-dimensional models for young SNRs arising from the 
interaction between Type Ia explosions with different \mch\ and sub-\mch\ progenitors 
and a uniform ambient medium. We have generated synthetic X-ray spectra for each model 
at different expansion ages, separating the reverse and forward shock contributions. 
Our model spectra are publicly available, and can easily be convolved with the spectral 
responses of current and future X-ray missions like \textit{Chandra}, \textit{XRISM}, 
and \textit{Athena}. We have studied the bulk spectral and dynamical properties of our 
models (Fe K$\alpha$ centroid energies and luminosities, radii, and expansion ages), and 
have found that they provide an excellent match to the observations of most known Type Ia 
SNRs, indicating that the majority of SN Ia progenitors do not seem to substantially modify 
their surroundings on scales of a few parsecs, at least in comparison with CC SN progenitors. 
In our models, the ambient medium density and expansion
age are the main contributors to the diversity of the bulk SNR properties, but detailed fits 
to X-ray spectra can discriminate progenitor properties. We have also identified a few 
objects that cannot be easily reproduced by SNR models with a uniform ambient medium 
interaction, notably RCW 86, which is known to be a cavity explosion, and N103B, which is 
probably interacting with dense material of some sort. A detailed exploration of the 
parameter space for CSM interaction in Type Ia SNRs is required to gain further insight from 
these objects.

\acknowledgments Support for this work has been provided by the Chandra Theory award TM8-19004X.
H.M.-R., C.B., and S.P. are funded by the NASA ADAP grant NNX15AM03G S01. 
H.M.-R. also acknowledges support from a PITT PACC and a Zaccheus Daniel Predoctoral
Fellowship. D.J.P. acknowledges support from the Chandra Theory Program NASA/TM6-17003X and the 
NASA contract NAS8-03060. S.-H.L. is supported by the Kyoto University Foundation 
(grant No. 203180500017). E.B. acknowledges funding from the MINECO-FEDER grant AYA2015-63588-P. 
The authors wish to thank the Lorentz Center and the organizers and participants
of the workshop ``Observational Signatures of Type Ia Supernova Progenitors (III)'' for
stimulating discussions that helped finish this work. We also thank Karin Sandstrom and 
Rachel Bezanson for assistance with references regarding the Galactic hydrogen density probability
distribution function. This research has made extensive use of NASA's Astrophysics Data 
System (ADS, \url{http://adswww.harvard.edu/}).

\software{\crcode\ \citep{Ell07,Pat09,Ell10,Pat10,Cas12,Lee12,Lee13,Lee14,Lee15}, \texttt{Matplotlib} 
\citep{Hun07}, \texttt{IPython} \citep{PeG07}, \texttt{Numpy} \citep{vaW11}, \texttt{AtomDB} 
\citep{Fo12, Fo14}, \texttt{PyAtomDB} (\url{http://atomdb.readthedocs.io/en/master/}), 
\texttt{Astropy} \citep{Astro13,Astro18}, \texttt{Python} (\url{https://www.python.org/}), 
\texttt{SciPy} (\url{https://www.scipy.org/}).}

\bibliographystyle{aasjournal}
\bibliography{Paper_SCH}

\end{document}